\documentclass{aa}

\usepackage{graphicx}
\usepackage{txfonts}
\usepackage{lipsum}
\usepackage{subcaption}         
\usepackage{lscape}             
\usepackage{placeins}           
\usepackage{comment}                               
\usepackage{hyperref}

\begin{document}

\title{The large-scale ordered magnetic field in the Galactic halo and the Local Bubble}

   \titlerunning{The ordered magnetic field in the halo}
   \authorrunning{Orlando}
   
 \subtitle{}

   \author{Elena Orlando\inst{}\thanks{email: orland.ele@gmail.com}
        }
        \institute{
University of Trieste, Department of Physics, 34127 Trieste, Italy\\
Stanford University, CA-94305, Stanford, USA\\
Eureka Scientific, CA-94602, Oakland, USA\\
The National Institute for Nuclear Physics (INFN), 34127 Trieste, Italy
}

  \abstract
   {All-sky synchrotron polarization observations are the primary probe of the Galactic large-scale ordered magnetic field, which includes both coherent and ordered random components. However, magnetic-field strength estimates from synchrotron observations depend on the assumed cosmic-ray electron model, which itself depends on the magnetic-field configuration and strength throughout the Galaxy.}
   {We aim to constrain the strength of the large-scale ordered magnetic field in the Galactic halo through a self-consistent treatment of cosmic-ray electrons with the latest coherent magnetic-field configuration inferred by Xu $\&$ Han from rotation measures corrected for local contamination.} 
  {We compare the {\it Planck} 30~GHz synchrotron polarization maps with our large-scale synchrotron models that jointly treat cosmic rays and magnetic fields and are  
  constrained by direct cosmic-ray measurements and multifrequency data. 
  Since the inferred field strength depends on the halo size, we explore different cosmic-ray propagation halo sizes.} 
   { We observe a significant large-scale synchrotron excess relative to model predictions, which requires an ordered toroidal magnetic-field component in the halo with a peak strength of \(3.2\text{--}4.3\,\mu\mathrm{G}\), depending on the assumed  halo size. We also find that the cosmic-ray propagation model with a 4~kpc halo size is preferred by the {\it Planck} synchrotron polarization maps.}
   {This result shows that the ordered magnetic field is 4–6 times stronger than the coherent field inferred from rotation measures by Xu $\&$ Han, establishing the presence of an ordered random field in the Galactic halo that is not traced by rotation measures.
   Attributing the synchrotron excess entirely to the Local Bubble would require non-standard Local Bubble properties, disfavoring this interpretation.}


   \maketitle
\nolinenumbers

\section{Introduction}
\label{intro}

On Galactic scales, the magnetic field is commonly described as the superposition of a coherent (also called regular) component, a turbulent (also called isotropic random) component, and a third component, commonly referred to as anisotropic random \citep{Beck2003}, ordered random \citep{Jaffe2010}, or striated \citep{JF2012}. 
Each component contributes differently to a variety of observables.
Briefly, all components contribute to the total synchrotron temperature.
Instead, the coherent can be traced through Faraday rotation and polarized synchrotron emission, while the ordered random can be traced by polarized synchrotron emission and it is not traced by Faraday rotations. Hence, while the configuration of the large-scale coherent magnetic field is usually obtained by rotation measures of pulsars or extragalactic sources, the sum of coherent and ordered random components, also called ordered component \citep{BeckRev1996}, is obtained from polarized synchrotron observations. 
Despite significant progress in characterizing the coherent and ordered magnetic field in the Galactic halo, the structure and strength of both components remain relatively uncertain. Comprehensive reviews include \citep{FerriereRev,Haverkorn,BeckRev,Han2017Rev,JaffeRev}.

The long-established optical starlight polarization \citep[e.g.][]{Davis,Heiles,Pelgrims,Gina} and polarized thermal dust emission \citep[e.g.][]{Crutcher2012,Planck2015a,DickinsonDust,Fauvet} have been widely used to successfully constrain the orientation of the ordered field, including in the Galactic halo, but do not directly probe the magnetic-field strength and are primarily sensitive to dust-rich regions.
A relatively recent approach uses the alignment of neutral hydrogen filaments to infer the orientation of the magnetic field \citep[e.g.][]{Clark}, including at high Galactic latitudes, providing high-resolution mapping of small to intermediate scale structures. 
Rotation measures of pulsars and extragalactic radio sources provide the primary probe of the large-scale coherent magnetic field in the halo, because they are sensitive to the parallel component of the magnetic field weighted by the thermal electron density. 
This method has long been used \citep[e.g.][]{Manchester,Han2004,Han2006} to trace the three-dimensional structure of the coherent magnetic field; however, constraints on its strength are influenced by uncertainties in the spatial distribution of thermal electrons.
Rotation measure distributions have suggested \citep{Han1997} the presence of a large-scale toroidal magnetic field structure in the  halo, which has opposite directions above and below the Galactic plane.
As more rotation measures become available, models of the coherent halo component, including toroidal and, in some cases, poloidal geometries, have become increasingly sophisticated, but no consensus has been reached on model parameters \citep[e.g.][]{Han1999,TT,Prouza,Sun2008,Taylor,Sun2010,Pshirkov,Mao,JF2012,FT2014,Xu2019,imagine,UF2024,Korochkin}. 
Moreover, the determination of the toroidal halo field is often strongly affected by the contamination of the rotation measures by small-scale local features. A notable recent study \citep{Xu2024} has obtained local-subtracted rotation measures after accounting for rotation measures produced by the local interstellar medium. These are used by the same authors to trace the large-scale coherent magnetic field free from local contamination. 
Another method of tracing the magnetic field in the halo is through polarized synchrotron emission  produced by Cosmic-Ray electrons (CRe) spiraling in the ordered magnetic field \citep[e.g.][]{Page,Miville,Jaffe2010,SOJ2011,Fauvet2012,JF2012,OS2013,PlanckBfield,O2019,UF2024}. Synchrotron polarization studies combined with rotation measures studies had suggested the presence of an ordered random component extending into the halo \citep{Jaffe2010,OS2013}. 
Polarized synchrotron observations provide high-quality, all-sky constraints on the ordered magnetic field component; however, its strength is degenerate with the assumed CRe usually taken as a fixed model or approximated to be in equipartition with the magnetic field. 
Indeed, the large-scale magnetic field affects the diffusion and energy losses of CR from their sources throughout the Galaxy, thereby influencing CR propagation and the spatial and spectral distribution of CRe across the Galaxy \citep[e.g.][and references therein]{StrongRev,O2018,O2019}.

Regarding synchrotron emission, a self-consistent, physically based treatment of the magnetic field in conjunction with CRe was first developed and applied by \citet{SOJ2011}, \citet{OS2013}, and \citet{O2018}, also using additional constraints from direct CR measurements and multiwavelength data, such as Fermi Large Area Telescope data \citep{Atwood}.
In the current work, following that approach, we obtain the strength of the large-scale ordered magnetic field in the halo using polarized synchrotron emission within CR propagation models, in which transport, energy losses, diffusion, and secondary CRe production depend on the assumed magnetic field model in each position throughout the Galaxy and are constrained by CR direct measurements and multifrequency data. 
We test two CR propagation models with different propagation halo sizes. We adopt the toroidal 3D magnetic field model of \citet{Xu2024}, which is corrected for local contamination.
Section 2 describes the method with the {\it Planck} data and our models, and Section 3 presents the results. The discussion and conclusions follow.

\section{Method}
In this section, we describe the {\it Planck} data maps used in our analysis and their preparation, the CR propagation models employed, and the methodology for comparing models with observations to ultimately determine the strength of the ordered magnetic field in the Galactic halo.

\subsection{Data} 
\label{Data}
We use the publicly available synchrotron polarization maps from {\it Planck} Public Data Release R3\footnote{\nolinkurl{COM\_CompMap\_QU-synchrotron-commander\_2048\_R3.00\_full.fits} from \nolinkurl{https://irsa.ipac.caltech.edu/}} \citep{PlanckDR3}, the B{\fontsize{8}{11}\selectfont EYOND}P{\fontsize{8}{11}\selectfont LANK}\footnote{\nolinkurl{BP\_synch\_IQU\_n1024\_v2.fits} from \nolinkurl{https://beyondplanck.science/products/files\_v2/}}  \citep{BPColl2023}, and the C{\fontsize{8}{11}\selectfont OSMOGLOBE }DR1\footnote{\nolinkurl{CG_synch\_IQU\_n1024\_v1.fits} from \nolinkurl{https://www.cosmoglobe.uio.no/products/cosmoglobe-dr1.html}} \citep{Watts2024} products. They are all used to test the robustness of our results. 
The {\it Planck} DR3 synchrotron maps include Stokes $Q$ and $U$ for the full mission. They have been obtained with the Bayesian framework \texttt{Commander} by modeling synchrotron with a spatially constant spectral index. The resulting maps contain lower contamination from instrumental systematic effects than in previous versions. 
Synchrotron maps have $N_{side}=2048$ following the \texttt{HEALPix} 
convention \citep{Healpix}.
The B{\fontsize{8}{11}\selectfont EYOND}P{\fontsize{8}{11}\selectfont LANK} synchrotron maps, as described in \citet{BPMap}, include Stokes $I$, $Q$, $U$, and their respective standard deviation (RMS, root-mean-square) maps. These maps account for both {\it Planck} and {\it WMAP} observations and have been obtained with the Bayesian framework and partitioning the sky into four large disjoint regions (High Latitude;
Galactic Spur; Galactic Plane; and Galactic Center), each associated with its own power-law index. Synchrotron maps have $N_{side}=1028$ following the \texttt{HEALPix} convention. 
The C{\fontsize{8}{11}\selectfont OSMOGLOBE }DR1 synchrotron maps, the framework of which is described in \citet{Watts2023a}, include Stokes $I$, $Q$, $U$, and their respective standard deviation maps, and have been derived similarly to the B{\fontsize{8}{11}\selectfont EYOND}P{\fontsize{8}{11}\selectfont LANK} maps from WMAP and Planck LFI frequency maps. They have significantly lower instrumental systematics than the legacy products from each experiment. Synchrotron maps have $N_{side}=1028$ following the \texttt{HEALPix} convention.

After testing that the three maps produce very similar results, in this work we show resulting maps using the second dataset only.

\begin{figure}[!ht]
   \centering
   \includegraphics[width=\hsize]{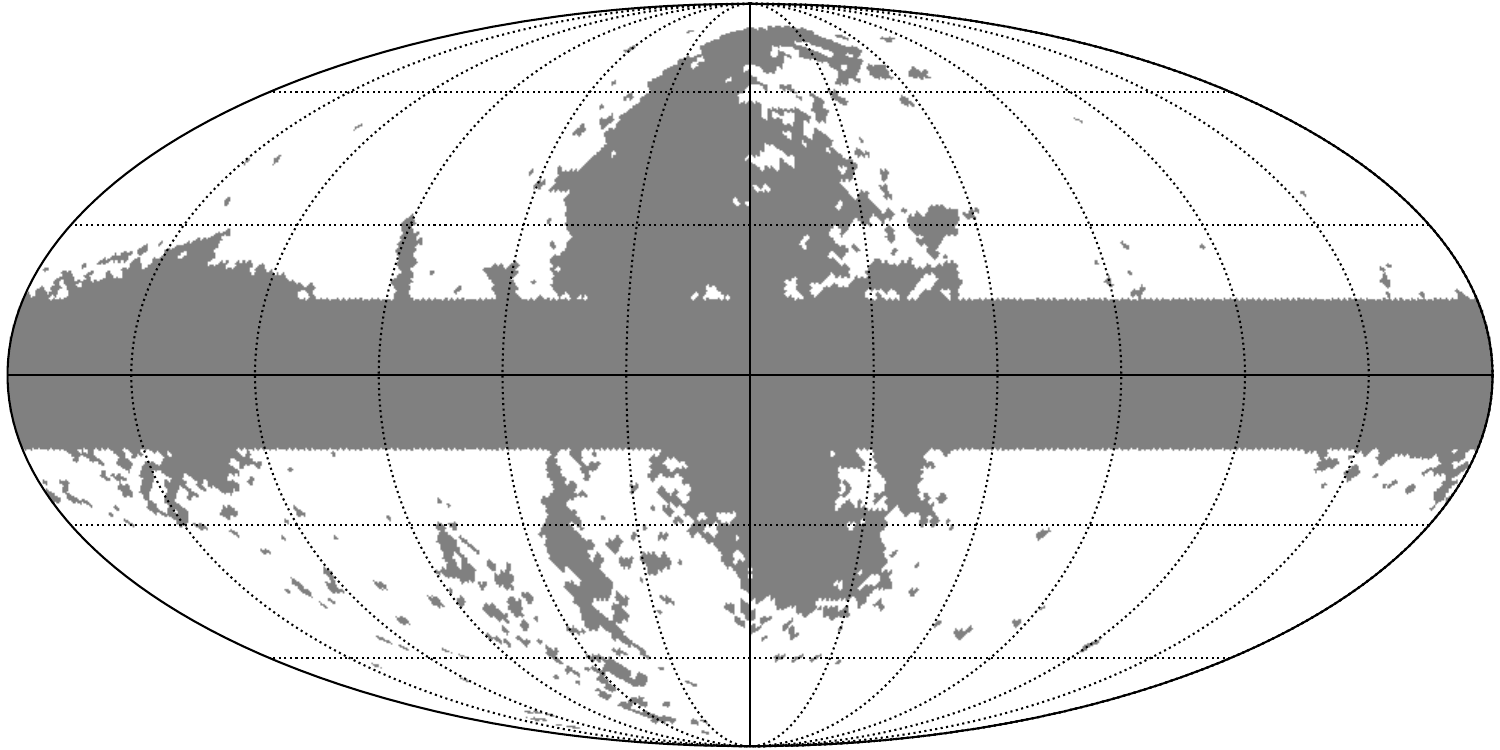}
      \caption{Mask used for this work.}
         \label{fig1}
   \end{figure}

As usual, the polarized intensity map $P$ is computed from $P = \sqrt{Q^{2} + U^{2}}$.
We smooth the {\it Planck} synchrotron maps with a 1$^\circ$ FWHM Gaussian kernel. Then we downgrade them to $N_{side}=64$ to match the pixel size to the smoothing scale and to optimize the signal-to-noise ratio for large-scale synchrotron polarization studies. This suppresses small-scale noise and improves signal-to-noise in faint regions, as commonly used in {\it Planck} analyses.
In addition, the downgrading ensures that the final map is consistent with the angular resolution of the synchrotron models.
We verify that the results are stable when adopting higher spatial resolutions and/or different smoothing scales, confirming that our results are not sensitive to the particular choice of smoothing and pixelization parameters. 
Uncertainties, as RMS, are treated in the same manner.
{\it Planck} polarization maps reveal a non-Gaussian distribution characterized by bright Galactic regions and loops. The latter are considered to be mainly local features \citep[e.g.][]{Vidal2015}. These bright local features are not present in our large-scale synchrotron model and neither in the magnetic field model, which was derived from rotation measures corrected for local contamination (see Section~\ref{bfield}). To remove the influence of these bright features while maintaining a representative fraction of the sky, we generate a binary mask that excludes the brightest 30\% of pixels, corresponding to the 70th percentile of the intensity distribution of the entire maps. 
The selected sky percentile is consistent with those reported in previous studies \citep[e.g.][]{PlanckDR3,Vidal2015,filaments}. 
Then, we construct an additional map, as described in the following. 
In the literature, there is no official latitude that defines the beginning of the halo. 
\citet{diffuse2} and \citet{OS2013} define intermediate Galactic latitudes as those between 
$10^\circ$ and $20^\circ$. 
\citet{Carretti}, using the Parkes Galactic Meridian Survey, pointed out that the strong disk emission in polarization extends to $|b| \approx 20^\circ$, while at higher latitudes the emission begins to decrease until the halo.
Because the halo magnetic field configuration (see Section~\ref{bfield}) reaches its maximum strength at a Galactocentric distance of $\approx$8~kpc and at a height of 3~kpc above the Galactic plane, which corresponds to a Galactic latitude of about $20^\circ$ toward the Galactic Center, we adopt a latitude cut of $\pm 15^\circ$ to ensure that our region encompasses the entire magnetic field halo. This mask also removes possible contamination of the magnetic field disk component, which will be studied elsewhere.
The resulting composite mask is shown in Fig.~\ref{fig1}.
For both data and models, we apply the mask to remove the unmodeled regions affected by strong emission and the Galactic plane.

\subsection{Synchrotron emission modeling}
\label{model}
Our theoretical synchrotron emission models are obtained with the extension in radio and microwaves of the GALPROP code\footnote{
\nolinkurl{https://galprop.stanford.edu} and\\
\nolinkurl{https://gitlab.mpcdf.mpg.de/aws/galprop}
}
as documented in \citet{OS2013}, where calculations of the Stokes parameters $I$, $Q$, and $U$ and the possibility of accounting for 3D formulations of the magnetic field were implemented in the code. In fact, \citet{SOJ2011} and \citet{OS2013} have extended GALPROP to include calculations of synchrotron temperature, synchrotron polarization, absorption, and free-free emission. 
GALPROP solves the transport equation for all required CR species \citep{SM98}. The propagation equation is solved numerically on a user-defined spatial grid, which in our case is in 3D, and energy grid. The solution proceeds until a steady-state solution is obtained from the heaviest primary CR nucleus to electrons and positrons. It takes into account diffusion, convection, energy losses, ionization, and diffusive reacceleration processes. Primary CR are injected, while secondary CR are produced by fragmentation and radioactive decay. Various propagation scenarios have recently been tested against the latest CR measurements \citep{SO2024}. 
As reported in \citet{OS2013}, we compute the synchrotron emissivities at every 3D grid point $(x,y,z)$, obtaining the local emissivity contributions to the Stokes $I$, $Q$, and $U$. We then integrate these emissivities along each line of sight to construct the corresponding synchrotron sky maps. 
Our method differs from earlier analytical approaches \citep[e.g][]{Hammurabi,Jaffe2013,JF2012} that relate the Stokes parameters directly to the components of the magnetic field, which are only valid under the assumption that the CRe spectrum is a pure power law. Following approaches \citep[e.g][]{PlanckBfield,UF2024} assume a predetermined CRe density distribution from e.g. \citet{SOJ2011}, \citet{OS2013} or \citet{O2018}.
Our formalism computes the emissivity with propagation and with the magnetic field used, providing a consistent description of the synchrotron emission.
We use a 3D spatial grid with steps of 0.1~kpc in the $z$ direction and 0.2~kpc in the $x$ and $y$ direction to account for the variations of the magnetic field model throughout the Galaxy. Like the data, the polarized intensity map $P$ is computed from $P = \sqrt{Q^{2} + U^{2}}$.

In general, for a typical interstellar magnetic field of 
$B~\simeq~(4$~--~$5)~\mu\mathrm{G}$, CRe that emit synchrotron radiation at 
$\nu \simeq 30~\mathrm{GHz}$ have a characteristic energy of 
$E \simeq 20~\mathrm{GeV}$. 
In more detail, the relation between the CRe energy $E_{\rm GeV}$ and the synchrotron emission frequency $\nu_{\rm GHz}$ follows from the standard expression for the critical synchrotron frequency \citep{RL}

\begin{equation}
\nu_c \simeq \frac{3}{2} \, \gamma^2 \frac{e B_\perp}{2 \pi m_e c}
\text{~~~that gives} \quad
E_{\rm GeV} \simeq 8 \left[ \frac{\nu_{\rm GHz}}{B_{\mu{\rm G}} \, \sin \alpha} \right]^{1/2}
\label{eq:residualP}
\end{equation}

where $B_\perp = B_{\mu{\rm G}} \, \sin \alpha$ is the perpendicular component of the magnetic field. 
The observed synchrotron intensity at 30~GHz depends not only on the magnetic field but also on the CRe density and spectrum. Consequently, our synchrotron emission model at 30~GHz, while dominated by CRe around 20~GeV, is affected by the spectral slope of the CRe on a larger energy band. Hence, variations in the CRe spectrum or magnetic field directly modify the synchrotron intensity and morphology \citep{SOJ2011,OS2013}.

Two baseline models based on our previous works \citep{O2018,O2019} are tested. Both have a pure diffusion scenario, as this was found to better reproduce spectral data from radio to gamma rays \citep{OS2013,O2018}, but they differ for the CR propagation halo size: 4~kpc and 10~kpc. The choice of these two values is motivated by the fact that the modeled halo magnetic field strength is more sensitive to the size of the CR propagation halo than to the other model parameters \citep{O2018}. Because our primary interest here is the strength of the halo magnetic field, this approach provides a reasonable way to bracket the true value. Note that the halo size is defined as the vertical extent of the region where CR diffuse and are confined, and it is implemented in our model as a boundary condition in the propagation equation, where the CR density is zero and the CR are free to escape from the Galaxy.
The details of the CRe and magnetic field modeling are described in the following subsections.

\subsubsection{CRe modeling}
In our modeling, the CRe spectrum and density are not assumed to be uniform throughout the Galaxy. Rather, they are modeled consistently during propagation and vary spatially, resulting in a physically motivated model.
Because the 3D configuration and strength of the magnetic fields affect CR propagation in the Galaxy and the associated energy losses, the resulting spatial distribution and spectrum of CRe depend directly on the assumed magnetic-field structure and strength. In our method, we account for this by using CR propagation models in a self-consistent way, modeling the 3D CR and magnetic field simultaneously.

The CRe that generate synchrotron emission at 30 GHz are not affected by solar modulation \citep[e.g.][]{AMS02a,Potgieter}, which bends the spectrum at lower energies. 
This makes the 30 GHz band robust against uncertainties in the CRe modulation, allowing it to be locally normalized to direct measurements. 
Indeed, the usual procedure with GALPROP is to obtain propagation parameters by fitting, after propagation, CR protons, helium, heavier nuclei, and their ratio to local measurements for a given CR source distribution, magnetic field configuration, interstellar photon density, and gas distribution.  
We model diffusion and energy losses self-consistently. 
The initial CR source distributions and propagation model parameters are based on the best-fit model found in \cite{O2018}. This best model (called PDDE in the original work) reproduces the local gamma-ray spectrum from the Fermi Large Area Telescope and the spectral shape of the radio and microwave data. 
This is a pure diffusion model with no reacceleration processes. The hadronic propagation parameters are taken as in \cite{O2018} and used in \cite{O2019}. 
Because the CRe spectrum in \cite{O2018} was derived using a different magnetic field model from the one adopted here, we first refit the CRe spectrum to the data to determine the corresponding model parameters. For consistency with propagation models, this procedure needs to be reiterated every time we change the model parameters, including the magnetic field strength.

\subsubsection{3D magnetic field model}
\label{bfield}
We have implemented a new 3D model in GALPROP. 
The magnetic field model has the same formulation as in \citet{Xu2024} for the disk and toroidal halo components, and as in \citet{FT2014} for the X-shape component.

In more detail, the model of the coherent magnetic field in the halo by \citep{Xu2024} is based on the local-subtracted rotation measures. These were obtained by the authors after discarding any rotation measure deviating from its 30 nearest neighbors by more than three standard deviations. They used the median to naturally mitigate the influence of outlier background rotation measures, yielding a robust estimate of the large-scale  contribution subtracted by local effects.
With this method, the rotation measures produced by local interstellar medium features within 3 kpc from the Sun are not included.
Our model implements their best fit large-scale toroidal halo field, which is modeled as an azimuthal component 
that changes sign across the Galactic plane and is confined both in 
radius and height. The vertical dependence is described by the 
exponential profile with the vertical scale height \(z_{\mathrm{0}}= 3~kpc\), while the radial dependence is described by a Gaussian toroid with the maximum field strength at \(R_{\mathrm{h}} = 7.97~kpc\) and a
scale radius of \(R_{\mathrm{t}}= 5.31~kpc\). The strength \(B_{H}\) 
was found by \citep{Xu2024} to be \(0.73 \pm 0.12\) $\mu$G based on rotation measures.
The resulting field is purely azimuthal  
and reverses 
symmetrically above and below the plane. This is

\begin{equation}
B_{\phi}(r,z) = 
\mathrm{sign}(z)\, B_{H}\,
\left( \frac{|z|}{z_{0}} \exp\!\left[-\frac{|z| - z_{0}}{z_{0}}\right] \right)
\exp\!\left[-\left(\frac{r - R_{\mathrm{h}}}{R_{\mathrm{t}}}\right)^{2}\right]
\label{eq2}
\end{equation}
with \(r = \sqrt{x^{2} + y^{2}}\) the cylindrical radius, which represents the Galactocentric radius and $|z|$ the vertical distance  from the Galactic plane. 

As commonly assumed \citep[e.g.][]{JF2012, OS2013, Jaffe2013, PlanckBfield, UF2024}, our ordered random component follows the same topology as the coherent field. In this case, the ordered random component is aligned with the coherent field and has the same formulation as in Eq.~\ref{eq2}, but its direction reverses randomly on small scales. This component contributes to the polarized synchrotron emission, but not to the rotation measures. Hence, the resulting ordered component can be written as \(B_{\phi}(r,z)\) in eq.~\ref{eq2} multiplied by a positive coefficient.

In addition, our model implements a poloidal X-shape component whose strength is lower than that of the toroidal halo component, ensuring that it does not affect the rotation measure calculations presented in \citep{Xu2024}.
Following model C in \citep{FT2014} (eq. 14 and 15), we assume the X-shape components in cylindrical coordinates $(r,\phi,z)$ as 

\begin{equation}
B_r = \left( \frac{2~a~r_1^3\, z}{r^2} \right) B_{z}(r_1),\\ 
B_\phi = 0,\\
B_z = \left( \frac{r_1^2}{r^2} \right) B_{z}(r_1)
\label{eq3}
\end{equation}
with the vertical radius \(r_1 = r/(1 + a~ z^{2})\), which decreases with increasing $|z|$, and
the vertical field component having a radial exponential variation of \(B_{z}(r_1) = B_X \exp\!\left(-{r_1}/{L}\right)\). As in \citet{FT2014} the assumed parameters are: the opening angle of the X-shape \(a = 0.33\), the radial decay of the field strength \(L = 2.1\)~kpc, the normalization strength of the X-shape component \(B_X\), which we set at 0.2~$\mu$G. 
Because \(B_r\) is proportional to $z$, it changes sign across the plane, producing symmetric upward and downward X-shape, while the  vertical component \(B_z\) dominates close to the symmetry axis and decreases rapidly with increasing $|z|$.  
By construction, the poloidal component has no azimuthal contribution.  
Consequently, this field does not significantly affect rotation-measure predictions, which are primarily sensitive to the line-of-sight component of the toroidal field.

While the focus of this work is on the ordered halo magnetic field, we also incorporate a coherent disk component and an isotropic random component to reliably model CR propagation and associated energy losses. 
The coherent disk model is taken from \citep{Xu2019,Xu2024} with the same formulation and fixed values from that work\footnote{This component will be studied more deeply elsewhere. However, we have checked that varying the normalization of this component does not affect the conclusions obtained in the present work.}. The isotropic random magnetic-field model is fixed to the best-fit as in \citet{O2018,O2019} with the following simple exponential law and values:
$B_{\mathrm{ran}}(r,z) = B_{0,\mathrm{ran}} 
\exp\!\left[-(r - R_{\odot})/R_{0,\mathrm{ran}}\right]
\exp\!\left(-|z|/z_{0,\mathrm{ran}}\right)$, 
where $B_{0,\mathrm{ran}} = 4.9~\mu\mathrm{G}$ sets its normalization at the solar position, and the scale lengths $R_{0,\mathrm{ran}} = 30~\mathrm{kpc}$ and $z_{0,\mathrm{ran}} = 4~\mathrm{kpc}$ determine how rapidly the field  decreases with increasing Galactocentric radius $r$ and vertical distance $|z|$ from the Galactic plane, respectively. The best-fit values were obtained in \citet{O2018,O2019} by fitting radio surveys, in particular the Haslam 408~Mhz map \citep{Haslam} reprocessed by \citet{408MHz}.

\subsection{Spatial template fit}
We estimate the total ordered field (the sum of the coherent and ordered random field) by letting the modeled \(B_{H}\) in Eq.~\ref{eq2} vary to spatially fit the {\it Planck} synchrotron polarization maps. A value of \(B_{H}\) larger than the original coherent field would indicate a synchrotron excess. We do not attempt to scan the full parameter space or construct new magnetic field models; hence, the magnetic field configuration and parameters are kept fixed as in \citet{Xu2024}. 
We produce \texttt{HEALPix} maps in polarization at 30 GHz for each model tested.
We check that, due to the smoothed spatial distribution of the synchrotron emission models in the halo, a model resolution of $N_{side}=64$ for the \texttt{HEALPix} map is reasonable, which corresponds to a resolution of approximately $1^\circ$. Hence, the model and the data have the same spatial resolution and number of pixels, and both are masked using the mask described in Section~\ref{Data}. The resulting model and data maps are then treated as templates for the fit.
Whenever a synchrotron model template is obtained for given magnetic field and CR propagation models, with the propagated CR spectrum at the solar position fitted to CR data, we perform the fit to estimate the best normalization value for the total ordered toroidal magnetic field in the halo. We define this value as \(B_{OH}\).
In more detail, initially, we compute the models using the original values for both $B_{H}$ and the injected CRe spectrum. We then determine the scaling factor of the original $B_{H}$ required for the model to reproduce the data by directly scaling the synchrotron model maps to roughly match the synchrotron data. Although the fitting could, in principle, be performed through multiple \texttt{GALPROP} runs with varying parameters, this is not necessary. This approach, also adopted by \citet{OS2013}, is faster, simpler, and yields the same results.
We estimate the best-fit scaling parameter $a$ by minimizing the $\chi^{2}$ statistic using the \texttt{iminuit} package \citep{iminuit}. The $\chi^{2}$ is defined as
\begin{equation}
\chi^{2} = \sum_i \frac{(D_i - a\, M_i - c)^2}{\sigma_i^2},
\label{eq4}
\end{equation}
where $i$ is the pixel index of the \texttt{HEALPix} maps, $D_i$ are the synchrotron polarization data, $M_i$ the synchrotron polarization model, and $\sigma_i$ the corresponding uncertainties. The constant $c$ is set to zero, consistent with standard \textit{Planck} polarization analyses. Only pixels with finite values are included in the fit. The free parameter $a$ is the multiplicative coefficient that rescales the model to the data by minimizing the $\chi^{2}$.
After fitting the polarized synchrotron normalization, the corresponding preliminary magnetic-field strength is obtained using the fact that $P \propto B_{H}^{2}$. Because this proportionality is strictly valid only for a CRe spectral index of 3, we subsequently re-run the models with the updated best-fit values for further refinement. In this second stage, the fitting is performed through multiple \texttt{GALPROP} runs and again using \texttt{iminuit} to minimize the $\chi^{2}$ and determine the best-fitting injection spectrum of primary CRe compatible with the given value of $B_{OH}$. In this second step, the synchrotron emission model needs to be produced for the best-fitting CRe injection spectrum. The entire procedure is iterated until the uncertainty in the scaling factor $a$ reaches the 3\% level.
This iterative strategy significantly reduces the computational load, which is substantial for full 3D synchrotron models.

\section{Results}
For each baseline model with \(z = 4~\mathrm{kpc}\)  and \(z = 10~\mathrm{kpc}\)  CR propagation halo size, this section includes the best-fit spectrum and distribution of CRe, the best-fit strength of the toroidal halo field, and the resulting synchrotron emission model maps at 30~GHz. The best synchrotron model is also identified. 

\subsection{CRe spectrum and distribution}
\begin{figure}[!ht]
   \centering
   \includegraphics[width=\hsize]{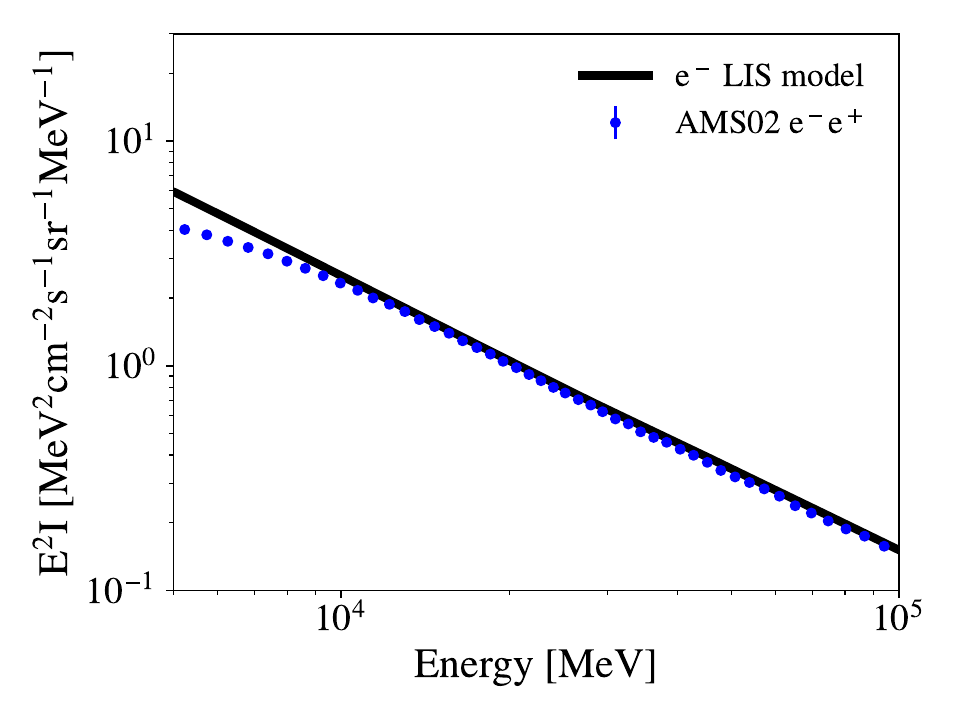}
      \caption{Propagated interstellar CRe spectrum of the best-fit model compared with AMS-02 CRe data from \citep{AMS02}.} 
         \label{fig2}
   \end{figure}

The resulting propagated CRe spectrum, known as the local interstellar spectrum (LIS), for the best-fit ordered magnetic field strength 
is shown in Fig.~\ref{fig2} for the energy interval centered around 20 GeV and calculated at the solar position. 
This is obtained following the usual procedure with GALPROP by fitting to the CR direct measurements the modeled local interstellar CR spectrum propagated at the solar position. The fit to the AMS-02 CRe data \citep{AMS02} is performed for the energy band (10~--~100) GeV. The normalization of CRe is \(1.27~\times~10^{-9}\,\mathrm{cm^{-2}\,sr^{-1}\,s^{-1}\,MeV^{-1}}\)
at 25 GeV for both models, while the best-fit injection spectra\footnote{The discrepancy between model and data below 10~GeV is due to solar modulation and it is not the aim of the study here since it does not affect the reported results.} are 2.55 for the model with \(z = 4\,\mathrm{kpc}\)
and 2.52 for the model with \(z = 10\,\mathrm{kpc}\). Following \citet{O2018} the diffusion coefficient is \(1.228~\times~10^{29}\,\mathrm{cm^{2}\,s^{-1}}\)
for the model with \(z = 4\,\mathrm{kpc}\) and \(2.136 \times 10^{29}\,\mathrm{cm^{2}\,s^{-1}}\) for the model with \(z = 10\,\mathrm{kpc}\). 
Other model parameters are fixed to the $PDDE$ model of \citet{O2018}. 
Because synchrotron radiation does not distinguish between emission from electrons and positrons, we treat both species as electrons in the model. 

   \begin{figure}[!ht]
   \centering
    \includegraphics[width=\hsize]{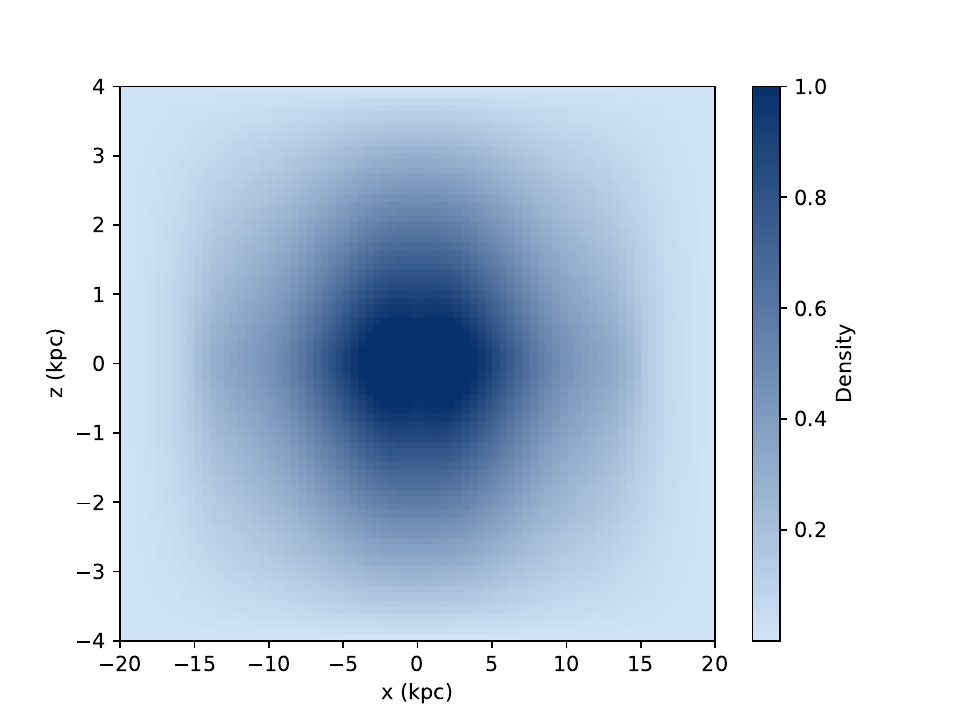}
   \includegraphics[width=\hsize]{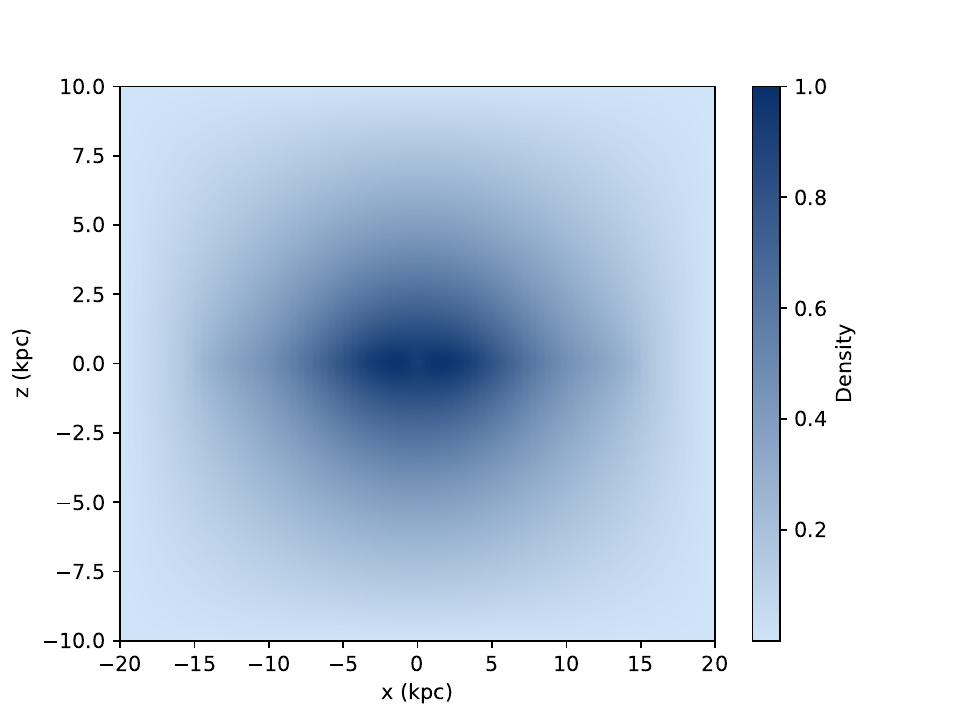}
      \caption{Spatial distribution of CRe at 20 GeV after propagation for the section through the 3D model, in x and z directions for y=0. Upper plot: model with halo size 4 kpc; lower plot: model with halo size 10 kpc. The density has arbitrary units and has the same normalization for both models. }
         \label{fig3}
   \end{figure}

Figure~\ref{fig3} shows the spatial distribution of CRe resulting after propagation for the best-fit models with \(z = 4~\mathrm{kpc}\)  and with \(z = 10~\mathrm{kpc}\) , evaluated in the $x$--$z$ plane at $y = 0$ for CRe energy of 20~GeV. Both density distributions reflect the spatial concentration of sources.
The difference in the 3D CRe distribution between the two models depends on their assumed CR propagation halo size, 
and their resulting magnetic field strength  
and diffusion coefficient. 
Hence, in GALPROP, the steady-state CRe distribution and spectra are determined by the interplay of CR injection spectra, spatial diffusion, and energy losses due to synchrotron and inverse Compton scattering. 
Overall, for the model with \(z = 10~\mathrm{kpc}\)  these effects result in a less centrally peaked distribution, with a more extended halo 
reflecting the combined effects of enhanced diffusion and decreased energy losses.

\subsection{The large-scale ordered magnetic field strength}
We observe a significant large-scale synchrotron excess relative to the model predictions using the original coherent component strength reported by \citet{Xu2024}. In fact, our modeled synchrotron emission does not match the overall normalization of the polarized data at 30 GHz, resulting in a best-fit magnetic field that is much larger than the coherent field in both models.
The original and best-fit strengths of the halo magnetic-field component are reported in Table~\ref{Table1} for each model with different halo sizes. We find that the resulting total ordered halo field, \(B_{OH}\), is \(4.3\,\mu\mathrm{G}\) for the model with \(z = 4~\mathrm{kpc}\) and \(3.2\,\mu\mathrm{G}\) for the model with \(z = 10~\mathrm{kpc}\). These values are approximately a factor of 4--6 larger than the original strength of the coherent field \(B_{H}\) inferred from the rotation measures. The factor of 4--6 is significantly larger than the expected  uncertainties in the strength of the coherent field (private communication with Prof. Han).
Because these synchrotron models represent the extreme values of the plausible values for the halo size, they can be used to bracket the total ordered halo field strength, which lies in the range \((3.2\text{--}4.3)\,\mu\mathrm{G}\).
The same Table also includes the $\chi^{2}$ values for the tabulated \(B_{H}\) and \(B_{OH}\). Compared with them, we obtain $\Delta \chi^{2}$
of 30513 and 29387 for the model with \(z = 4~\mathrm{kpc}\) and for the model with \(z = 10~\mathrm{kpc}\), respectively. 
Since fitting the model introduces an additional free parameter (a global normalization), the original and the fitted models are nested and $\Delta\chi^{2}$ follows a $\chi^{2}$ distribution with one degree of freedom. The large $\Delta\chi^{2}$ values statistically show that the  \(B_{OH}\) values are required by the data. 
As usually done \citep[e.g.][]{OS2013, PlanckBfield, diffuse2}, we use $\chi^{2}$ strictly as a comparative metric between models, since both fits are performed on the same dataset and with the same method. In this context, differences in $\chi^{2}$ directly reflect relative model performance, independent of the effective number of degrees of freedom.
Although the absolute value of the reduced $\chi^{2}$ is not interpreted as a strict goodness-of-fit measure, the $\Delta\chi^{2}$ comparison between nested models remains valid. The improvement is large enough to claim a synchrotron excess and make the inclusion of an ordered random component statistically indisputable. 
Since both models have the same number of free parameters, the $\chi^{2}$ provides a robust model comparison even when its absolute value is not used to access the goodness of the fit, as in this case. Hence, the model with the smaller $\chi^{2}$ is statistically preferred, which corresponds to the model with \(z = 4~\mathrm{kpc}\).

\begin{table}[h!]
\caption{Original coherent value and best-fit ordered values for the toroidal halo field}               
\label{Table1}    
\centering                        
\begin{tabular}{c c c}      
\hline\hline               
  ~~~~~~~~ ~~~ ~~~~   & model  & model  \\     
  ~~~~~~~~ ~~~ ~~~~   &  z~=~4 kpc &  z~=~10 kpc \\ 
\hline                      
\(B_{H}\) ($\mu$G) &  &  \\   
coherent only, &  & \\
original value  & 0.73 & 0.73  \\ 
from rotation measures  &  &  \\ 
\citep{Xu2024} & & \\
\hline   
$\chi^{2}_{B_{H}}$ & 40097   & 40731 \\
\hline  
\(B_{OH}\) ($\mu$G) &  &  \\ 
total ordered, & 4.3 & 3.2 \\
from synchrotron & & \\
(this work) & & \\
\hline   
$\chi^{2}_{B_{OH}}$ & 9584   & 11344 \\
\hline                                 
\end{tabular}
\end{table}

\subsection{Synchrotron model maps and residuals}
\begin{figure*}[!ht]
   \centering
   \includegraphics[width=172pt]{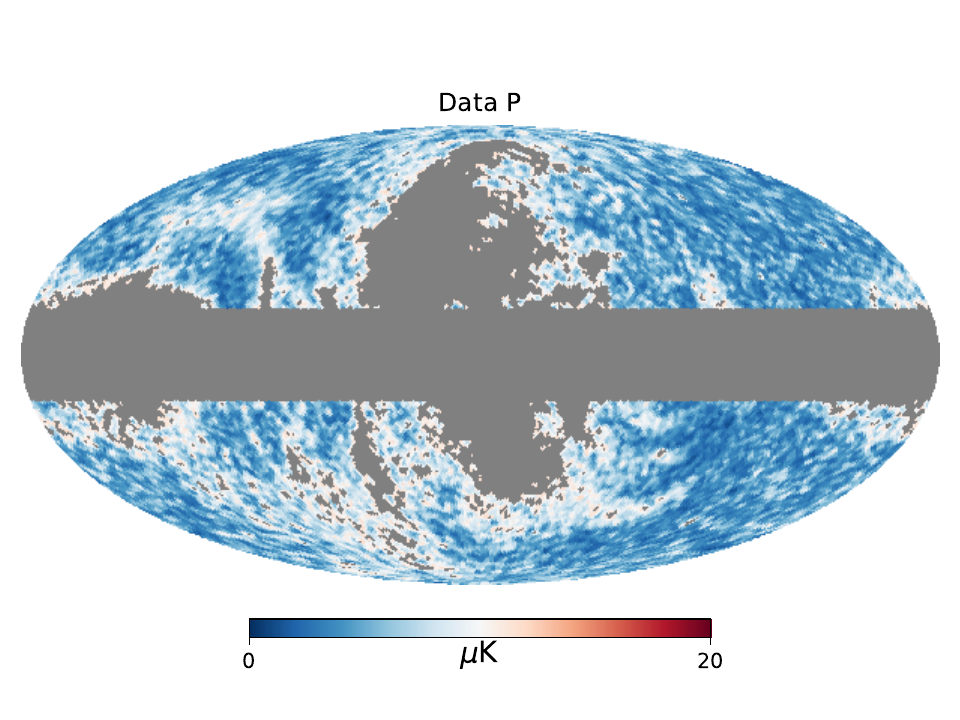}
      \includegraphics[width=172pt]{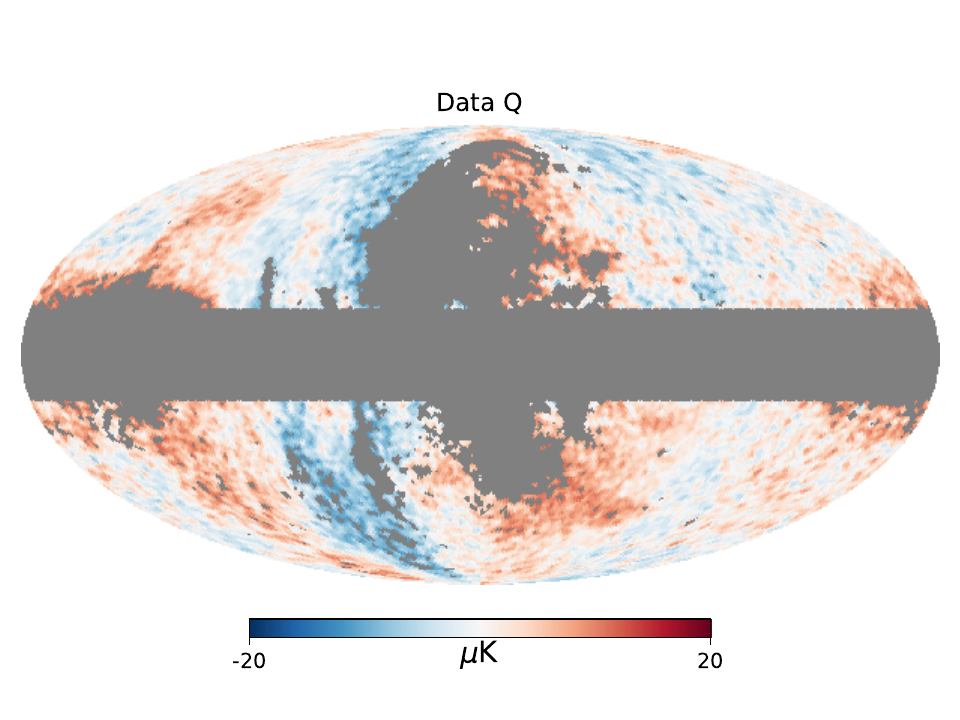}
            \includegraphics[width=172pt]{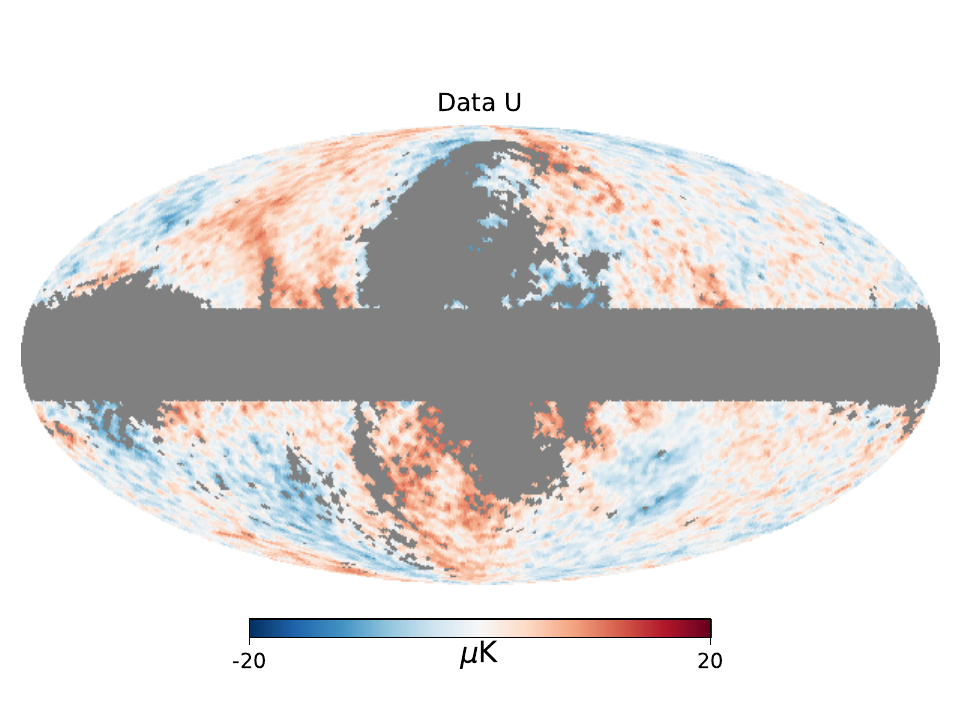}\\
   \includegraphics[width=172pt]{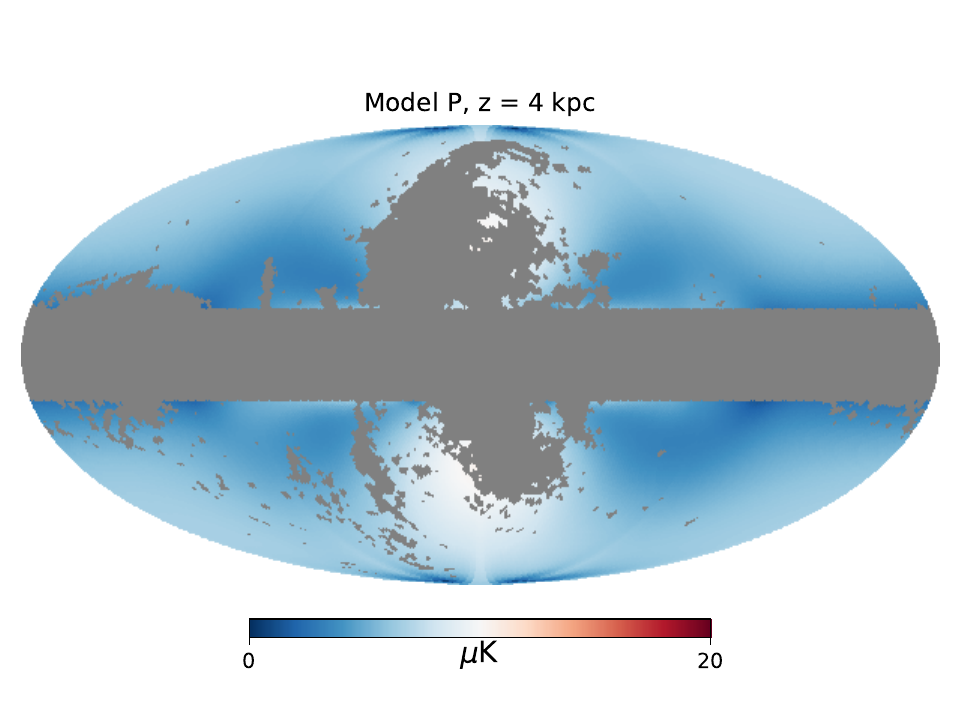}
      \includegraphics[width=172pt]{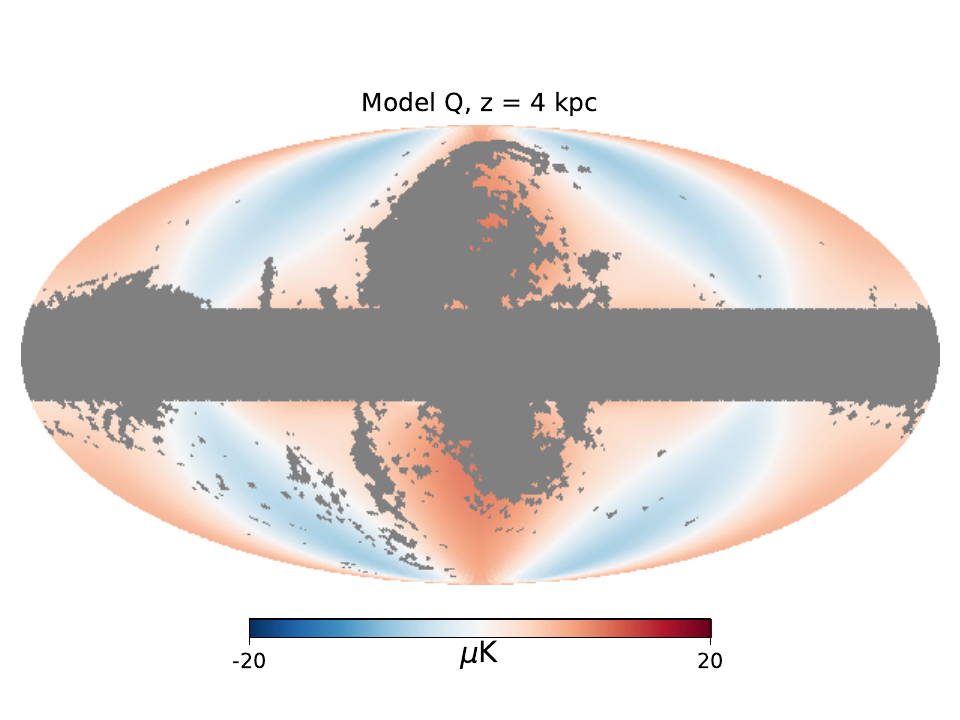}
            \includegraphics[width=172pt]{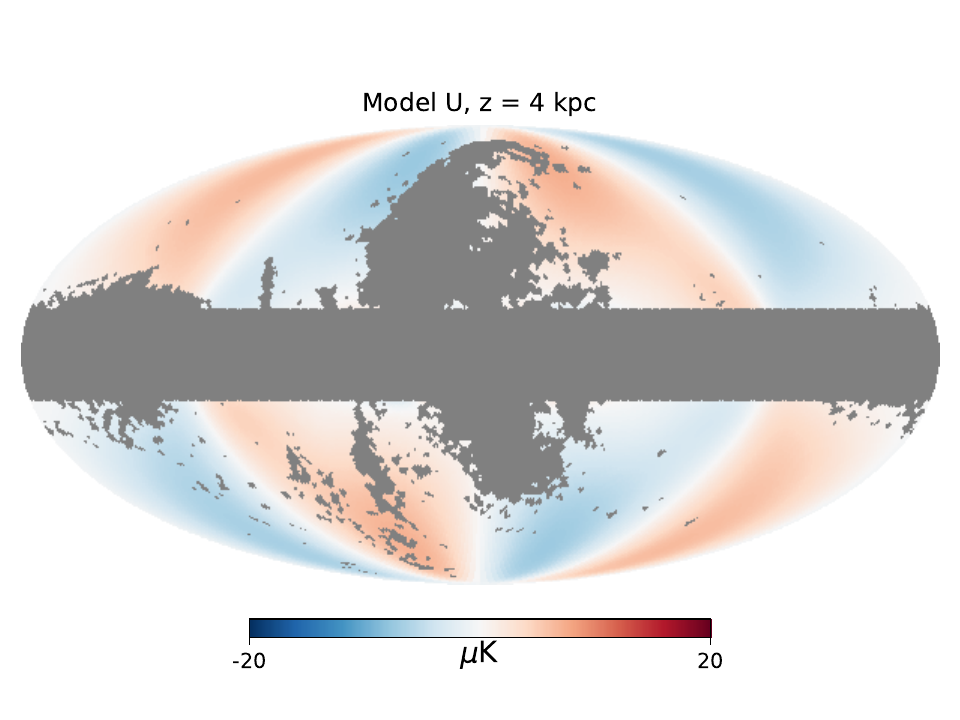}\\
               \includegraphics[width=172pt]{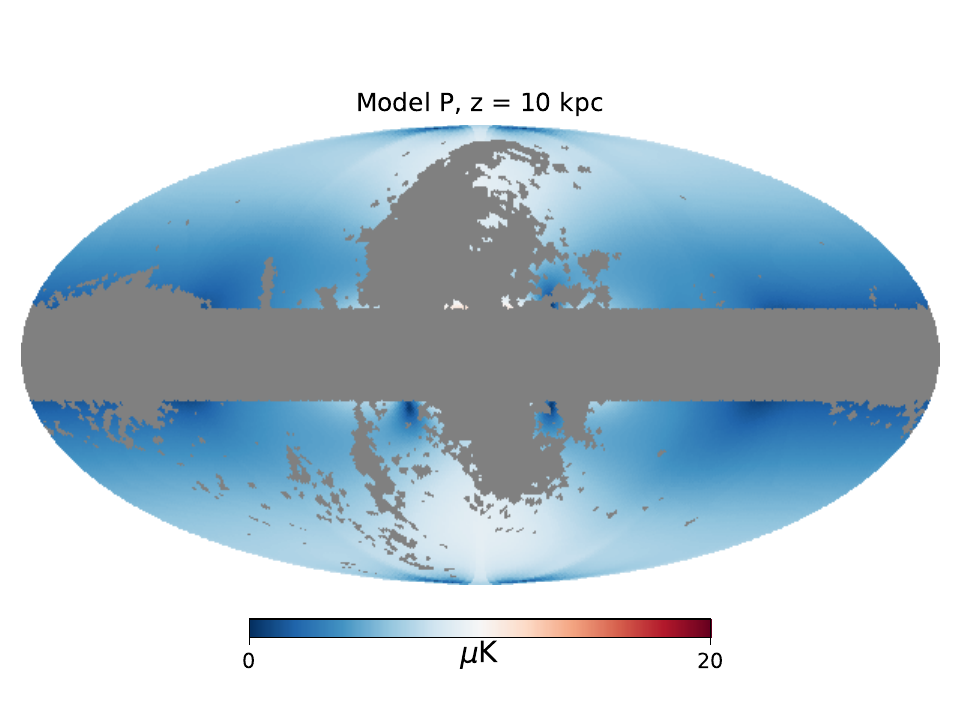}
      \includegraphics[width=172pt]{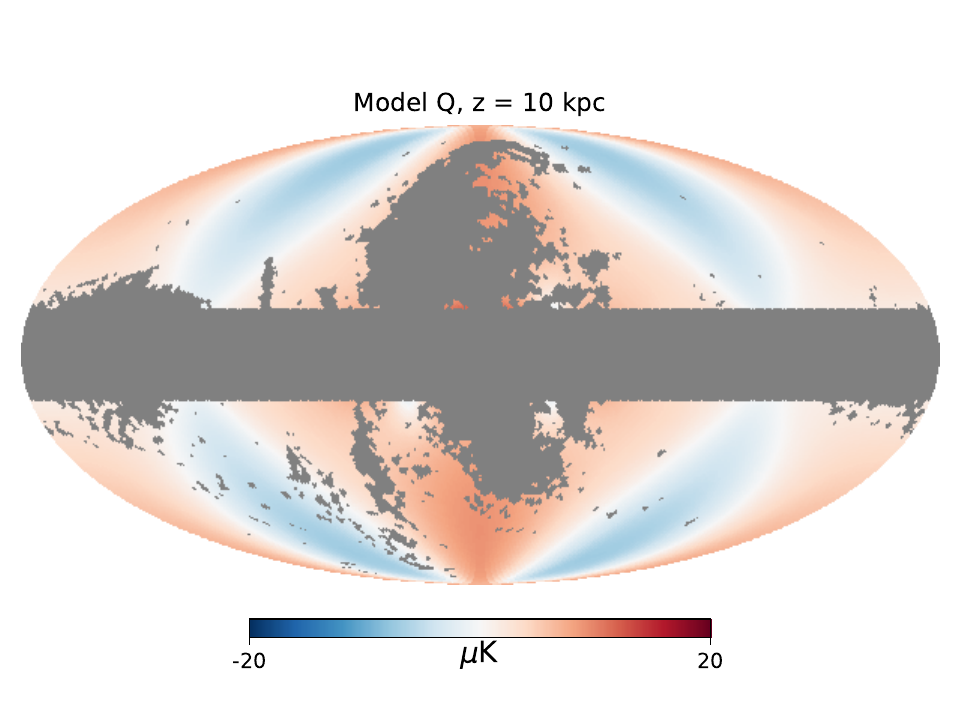}
            \includegraphics[width=172pt]{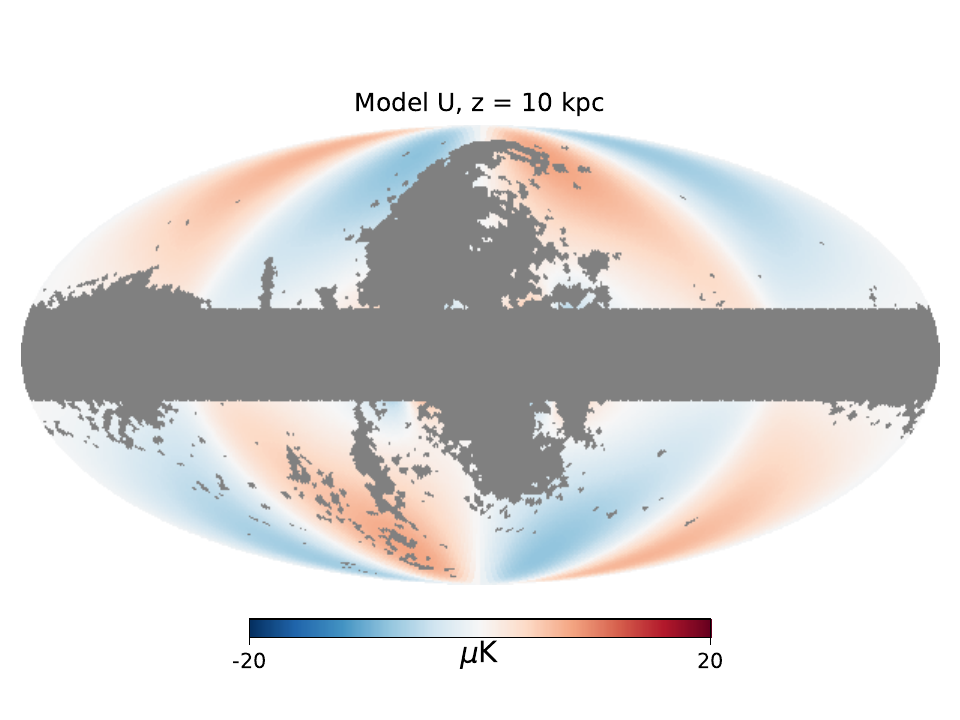}\\
      \caption{{\it Planck} synchrotron maps (top row) and our all-sky synchrotron prediction maps for both best-fit models with \(z = 4~\mathrm{kpc}\) (second row) and \(z = 10~\mathrm{kpc}\) (bottom row) in Mollweide projection and at 30~GHz. Left to right: polarized intensity $P$, Stokes $Q$, and Stokes $U$. The colorscale is the same for Stokes $Q$ and $U$, while is positive only for P. The gray region is masked.}
         \label{fig4}
   \end{figure*}

Figure~\ref{fig4} shows all-sky {\it Planck} maps (top row) and our all-sky model maps for both models with \(z = 4~\mathrm{kpc}\) (second row) and \(z = 10~\mathrm{kpc}\) (bottom row). The models are generated by using their best-fit values for the ordered halo magnetic field and CRe. Polarized intensity maps $P$, Stokes $Q$, and Stokes $U$ are shown for 30~GHz in Mollweide projection, with longitude increasing eastward (to the left) and latitude increasing northward (upward). The updated models broadly reproduce the main large-scale morphology of the synchrotron polarized intensity of the unmasked regions observed by {\it Planck}. The Stokes $Q$ $U$ model maps show consistency with the global topology of the magnetic field, although they are still in tension in some regions.   

\begin{figure}[!ht]
   \centering
      \includegraphics[width=260pt]{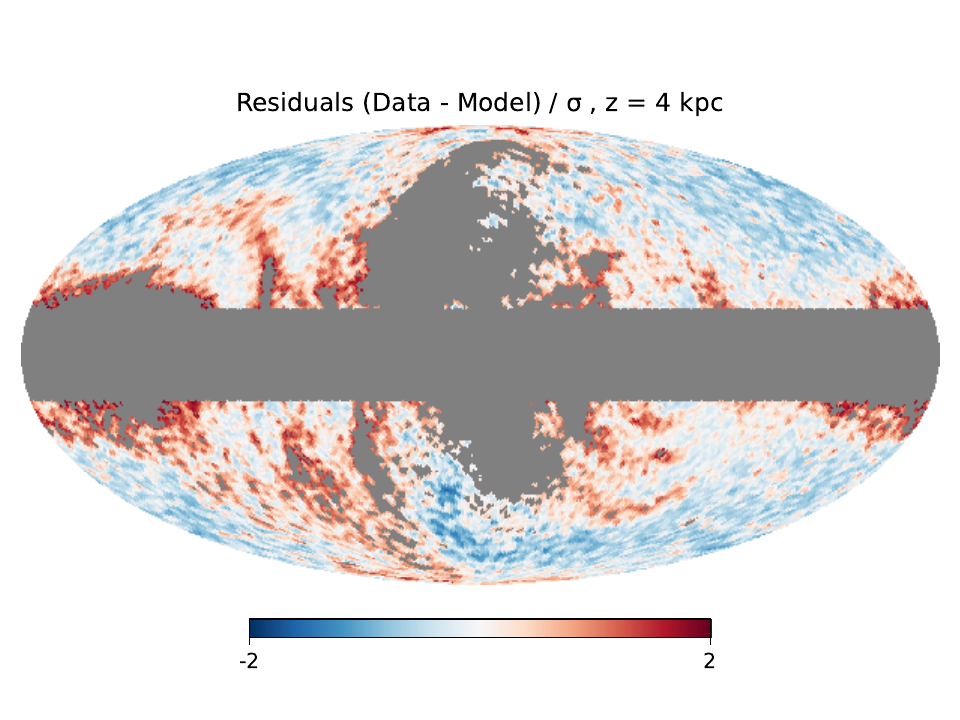}
            \includegraphics[width=260pt]{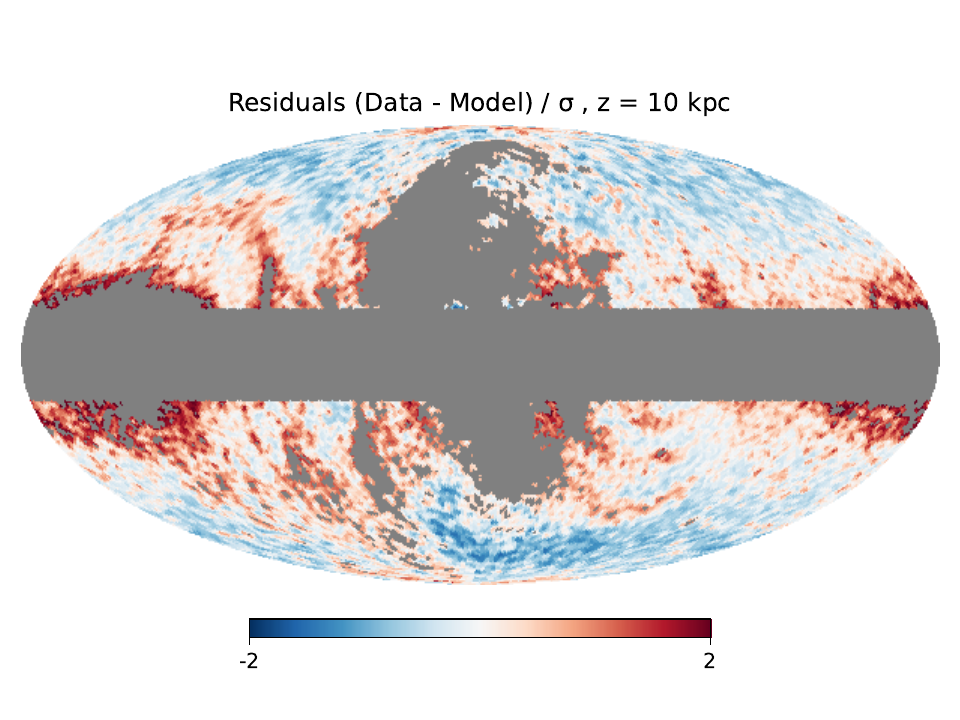}
      \caption{Polarized residual maps for the two best-fit models with \(z = 4~\mathrm{kpc}\) (top map) and \(z = 10~\mathrm{kpc}\) (bottom map).} 
         \label{fig5}
   \end{figure}

\begin{figure}[!ht]
   \centering
      \includegraphics[width=260pt]{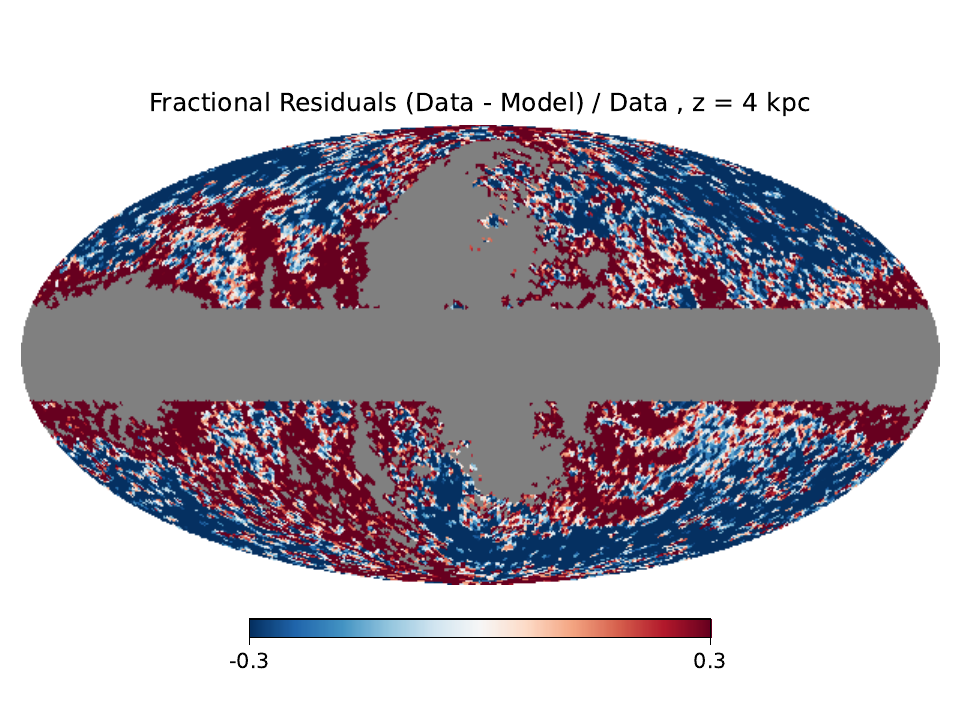}
            \includegraphics[width=260pt]{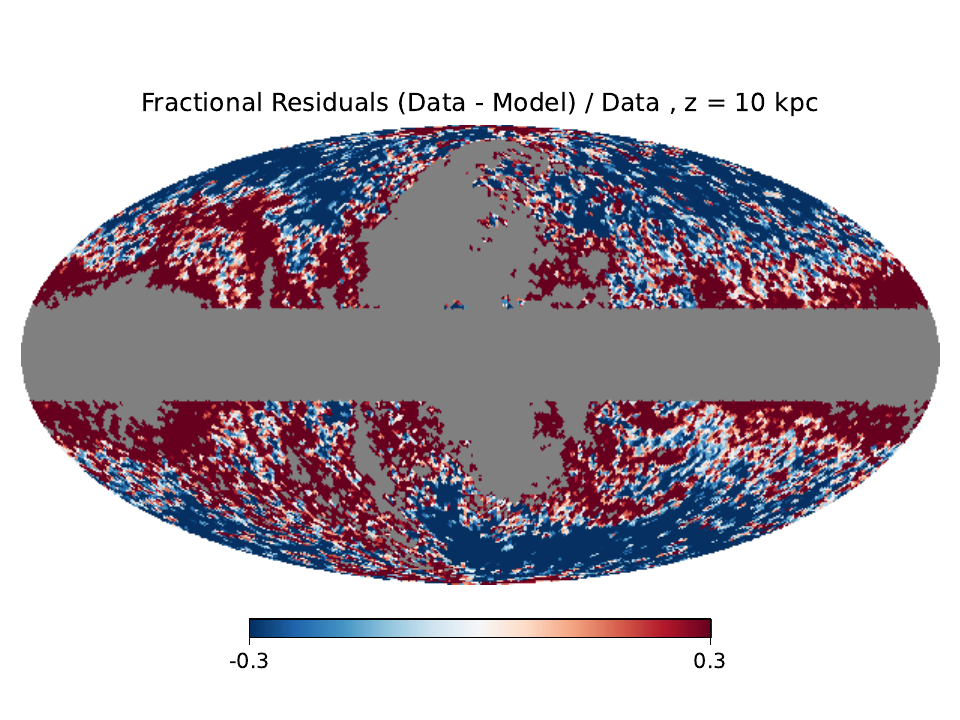}
      \caption{Polarized fractional residual maps for the two best-fit models with \(z = 4~\mathrm{kpc}\) (top map) and \(z = 10~\mathrm{kpc}\) (bottom map).} 
         \label{fig6}
   \end{figure}

To better appreciate the differences between best-fit models and data, we calculate the residual maps in polarized intensity as pixel-by-pixel differences between data and model by accounting for the RMS noise level (reported as $\sigma$), which provides a measure of errors. Moreover, an alternative and often more interpretable way of comparing the models with the data is through fractional residual maps. Residuals and fractional residuals of the polarized intensity are obtained as
\begin{equation}
\begin{aligned}
&\mathrm{Residual}_{Pi} =
\frac{D_{Pi} - M_{Pi}}
{\sqrt{\sigma_{Qi}^{2} + \sigma_{Ui}^{2}}} =  \frac{D_{Pi} - M_{Pi}}
{\sigma_{Pi}} \\[6pt]
&\mathrm{Fractional\ Residual}_{Pi} =
\frac{D_{Pi} - M_{i}}{D_{Pi}}
\end{aligned}
\label{eq5}
\end{equation}
\text{with} \quad
$D_{Pi} = \sqrt{D_{Qi}^{2} + D_{Ui}^{2}}$  polarized intensity of the data in pixel $i$
\quad
and $M_{Pi} = \sqrt{M_{Qi}^{2} + M_{Ui}^{2}}$ polarized intensity of the model in pixel $i$.
$\sigma_{Qi}$ and $\sigma_{Qi}$ are the RMS as reported in the original data file.
Residual maps are shown in Fig.~\ref{fig5} for the two best-fit models with \(z = 4~\mathrm{kpc}\) (top map) and \(z = 10~\mathrm{kpc}\) (bottom map) with extreme values of $\pm$2 to enhance contrast. The fractional residuals are shown in Fig.~\ref{fig6} for the same models. 
Both models exhibit large-scale residuals with similar features. The residuals show a mixture of positive and negative regions. Positive residuals, where the models underestimate data, occur primarily near the masked regions associated with the Fan and anti-center area, while both positive and negative residuals are more broadly distributed across the rest of the map. Some positive fractional residuals may resemble some filamentary structures as derived by \citet{filaments}.
In some regions, fractional residuals exceed $\pm$30\%. 
There are large-scale positive and negative differences between the models that are distributed over the entire sky. 
Overall, the model with the larger halo size increases the percentage of fractional residuals. This likely explains why the model with the smaller halo size is favored by the $\chi^{2}$ statistic; this 4~kpc-halo model is also the one conventionally used in gamma-ray and CR studies.

\subsection{Can the Local Bubble entirely explain the large-scale synchrotron excess?}
In this section, we analyze the possibility that the Local Bubble can entirely explain the large-scale synchrotron excess without invoking an ordered random component in the halo, as proposed by \citet{Korochkin} and \citet{Pelgrims2025}.

Being a region characterized by low gas density, a presumably low magnetic field, and a low CR density due to either fast diffusion or shielding by the bubble itself, 
accurate calculations of the Local Bubble's contribution to the polarized synchrotron emission would require a dedicated numerical modeling with very fine resolution, 
beyond the scope of the present study.  
However, we can provide analytical estimates of the expected synchrotron emission by adopting a geometrical model of the Local Bubble 
based on \citet{Korochkin}, \citet{Alves}, and \citet{Pelgrims2020,Pelgrims2025}, treating it as a spherical cavity with an inner radius of about 200~pc and a compressed magnetic field in a concentric shell. 
Assuming the magnetic field inside the Local Bubble is negligible, as in \citet{Korochkin} and \citet{Pelgrims2025}, 
the synchrotron emission from this inner region would be zero, 
while the emission from the shell could be significant if the magnetic field is largely compressed relative to the average Galactic field. 
In the following we analytically calculate the compression factor of the disk magnetic field within the Local Bubble that would be needed to compensate for the synchrotron intensity of the ordered random halo component.
We note that the synchrotron emission integrated along the line of sight does not depend on the proximity of the emitting region. Considering the most conservative scenario for the line of sight from the position of the Sun along the +z direction, the synchrotron intensity produced by the best-fit ordered random component integrated along the full line of sight is:

\begin{equation}
I_{OR} \propto \int_0^{4kpc} n_{CRe}~\left( 4.9~B_\perp \right)^2 ~dz
\label{eq6}
\end{equation}

The factor 4.9 is the result of $(B_{OH}-B_{H})/B_{H}$ reported in Table~\ref{Table1} for the halo size model of 4~kpc and only accounts for the ordered random magnetic field component. $B_\perp$ is the perpendicular component of the toroidal magnetic field in the halo. Our resulting $z$ distribution of CR electrons for the model of a 4~kpc halo at the solar position can be described by the following stretched exponential:

\begin{equation}
n_{CRe} \propto \,\exp\left[-\left(\frac{|z|}{2.5}\right)^{2}\right]
\label{eq7}
\end{equation}

Substituting eq.~\ref{eq7} and eq.~\ref{eq2}, eq.~\ref{eq6} becomes

\begin{equation}
\begin{aligned}
I_{OR} & \propto \int_0^{4kpc} e^{-\left(|z|/2.5\right)^2} ~\times \\ 
&  ~~~~~~~~~~~~~~~~~~~\left[ 4.9~ B_{H}\
\left( \frac{|z|}{z_{0}}~e^{-\frac{|z| - z_{0}}{z_{0}}} \right)~e^{-\left(\frac{r - R_{\mathrm{h}}}{R_{\mathrm{t}}}\right)^{2}} \right]^2 dz~=~14.1
\end{aligned}
\label{eq8}
\end{equation}

The average strength of the magnetic field in the region occupied by the Local Bubble in the $+z$ direction is obtained by :

\begin{equation}
\begin{aligned}
\langle B_{LB} \rangle & =
 \int_{0}^{0.2kpc} B_{disk} ~~dz~~\Big/
\int_{0}^{0.2kpc} dz \\
& = \frac{1}{0.2} \int_0^{0.2kpc} B_0~~e^{-R_G/A}~~ e^{-|z|/z_H}~~dz~~ \\
& = 5 \int_0^{0.2kpc} 1.65~~e^{-|z|/0.4}~~dz = 1.3\\
\end{aligned}
\label{eq9}
\end{equation}

where $B_{disk}$ is the coherent disk model by \citep{Xu2019,Xu2024} at the solar position  with the same formulation and the fixed parameter values from that work.

In order for the synchrotron in the shell of the Local Bubble, $I_{SHELL}$, to fully compensate for the synchrotron produced by the ordered random component, $I_{OR}$ in eq.~\ref{eq6}, the magnetic field in the shell $B_{SHELL}$ has to be compressed with a magnetic field compression factor $b~=~B_{SHELL}~/~\langle B_{LB} \rangle~$, which has to satisfy the following equation:

\begin{equation}
\begin{aligned}
\frac{I_{OR}}{I_{SHELL}}~=~1
& ~= 14.1~\Big/ \int_x^{0.2kpc} e^{-\left(0.2/2.5\right)^2}~B_{SHELL}^2 ~~dz~=\\
& ~= 14.1~\Big/ \int_x^{0.2kpc} ~\left[ b~\langle B_{LB} \rangle \right] ^2 ~~dz~=\\
&~=~14.1~\Big/ [1.7~b^2~(0.2-x) ]~=~41.5/b 
\end{aligned}
\label{eq10}
\end{equation}

with $(0.2-x)=h$, the thickness of the shell in kpc, which can also be written as $h~=~0.2/b$~kpc. From this, the magnetic compression factor becomes $b=41.5$. For a 200~pc bubble, this factor results  in a thickness of the Local Bubble's shell $h\approx 5pc$ and a $B_{SHELL}\approx 54~\mu G$. The CRe density at a distance of 200 pc is proportional to $e^{-(0.2/2.5)^2}$. These calculations assume that CRe penetrate the shell efficiently. For a Local Bubble model of extension 230~pc, as in \citet{Korochkin} and \citet{Pelgrims2025}, the same procedure results in $h\approx 6~pc$ and $B_{SHELL}\approx 49~\mu G$. Recalculating eq.~\ref{eq9} and eq.\ref{eq10} for a larger bubble would require a magnetic compression factor $b=33$ in a Local Bubble model of extension 500~pc with a $B_{SHELL}\approx 31~\mu G$ in a shell thickness $h\approx 15pc$. We have verified that using the less favored halo-size model of $10~\mathrm{kpc}$ produces similar results, as the lower ordered random magnetic-field component required to fit the synchrotron excess is compensated by the longer integration path and the different CR distribution.

\section{Discussion and conclusions}
While rotation measures trace the coherent component of the magnetic field, synchrotron polarization traces the total ordered component. 
We have used the {\it Planck} 30 GHz synchrotron polarization maps, which are not affected by Faraday rotation, they are free of contamination by polarized dust, and with depolarization negligible \citep{Page,Bennett,Planck2016}. 
This has allowed us to constrain the strength of the ordered toroidal magnetic field component in the halo, adopting the configuration of the magnetic field obtained by \citet{Xu2024} from rotation measures, which are corrected for local contamination. 

\paragraph{The X-shape field.} 
With respect to the original model in \citet{Xu2019,Xu2024} we have also included a dipolar field and, more specifically, an X-shape field as supported by various studies. 
In fact, in a rotating galaxy such as the Milky Way, the coexistence of a strong toroidal halo field with a weaker poloidal component follows naturally from dynamo theory \citep[e.g][]{Parker71, Brandenburg, FT2014,BeckRev,FerriereDynamo,FerriereNature,Hanasz}. 
Consequently, even though the toroidal halo field typically dominates in strength, a non-zero poloidal component must coexist to sustain the dynamo.
The model of the X-shape field is taken from \citet{FT2014} with a lower maximum strength than the toroidal halo field. 
Its strength would be better investigated when studying the polarized synchrotron emission in Galactic center regions that have been masked in this work. This modeled X-shape does not significantly affect rotation-measure predictions, which are primarily sensitive to the line-of-sight component of toroidal fields, nor synchrotron emission predictions due to its relatively low strength in the halo. 

\paragraph{The propagation halo size.} 
CRe are strongly affected by the assumed magnetic field model, but also by variations in the diffusion coefficient and halo size due to their rapid energy losses. Therefore, here, we have modeled them consistently alongside the magnetic field with CR propagation models. The halo magnetic-field strength and the CRe propagation halo size are degenerate \citep{OS2013}. More specifically, increasing the halo size has the strongest effect in producing a higher overall CRe density in the halo, which in turn leads to a lower inferred magnetic field strength in this region. To account for this uncertainty, we have tested two CR propagation models differing in the CR propagation halo size (\(z = 4~\mathrm{kpc}\) and \(z = 10~\mathrm{kpc}\)), values that are constrained by radioactive isotope data \citep[e.g.][]{StrongRev} and in agreement with observations in gamma rays \citep{diffuse2} and in radio from edge-on external galaxies \citep{Stein}. The choice of using such extreme values of the CR propagation halo size is motivated by the  sensitivity of the inferred halo magnetic-field strength to the halo size \citep{OS2013}, making these models ideal for bracketing its plausible range.
Moreover, local CR data mainly constrain the ratio $z^2 / D_0$, with $D_0$ diffusion coefficient. This leads to a degeneracy: models that reproduce the same local CR measurements have $D_0 \propto z$. Synchrotron emission, however, depends on the CR spatial distribution and therefore provides additional constraints that can partially break this degeneracy. 
Based on synchrotron polarization data in the halo, here we find that the CR propagation model with a halo size of 4~kpc is strongly preferred over the model with a 10~kpc halo. 

\paragraph{The strength of the ordered random field.} 
After computing synchrotron polarization maps using the original value of the toroidal halo magnetic field from \citet{Xu2024}, derived from rotation measures cleaned of local effects, we find that the predicted intensity significantly underestimates the intensity observed in the {\it Planck} 30~GHz map for both halo size models.
This implies that either the strength of the halo magnetic field reported by \citet{Xu2024} is underestimated, or an additional ordered random component must be present in the halo. Given the large discrepancy relative to the original values and the accurate work in \citep{Xu2024}, we consider the latter hypothesis to be the most plausible explanation. Moreover, we estimated that the Local Bubble alone would not be able to explain such a large discrepancy without invoking a large-scale ordered random component. 

\paragraph{Residual structures.} 
When comparing polarized synchrotron maps of models with {\it Planck} data, small-amplitude residuals of both signs appear in different regions of the sky, a behavior that has been observed in previous studies of microwave \citep{OS2013} and gamma-ray \citep{diffuse2} data. 
In general, both best-fit models exhibit good agreement with the 30~GHz {\it Planck} synchrotron polarization maps, despite the minimal fitting of magnetic-field parameters to synchrotron data. 
Our models match the observations within approximately 30\% across most of the sky, with deviations in specific regions discussed below. Many regions with enhanced emission were already underlined in previous studies \citep[e.g.][]{Jaffe2013,OS2013,PlanckBfield,Quijote,UF2024}. 
In these studies, numerous magnetic field model parameters were tuned using synchrotron maps as well; however, the resulting residuals did not show improvements relative to our results. Consequently, it is noteworthy that we achieve comparable agreement using only a simple fit of the overall normalization of a magnetic field model derived from rotation-measure data alone.
Also, given the limited flexibility in fitting the magnetic field model parameters to the 30~GHz data, we are not attempting a detailed characterization of the residual features. 
Although the best-fit models generally agree well with the data, residual maps reveal structures both on large and small scales. Small-scale discrepancies are an intrinsic limitation of large-scale GALPROP based models, arising from the necessary simplifications in the simultaneous modeling of CR transport, source populations, and magnetic-field distributions. In addition, some structures may be influenced by unmodeled loops and filaments \citep[e.g.][]{MS,Vidal2015,Planck2016,DickinsonLoops,Alves,Guidi}.
The most prominent large-scale residual is the excess observed in the outer Galaxy. Specifically, the anti-center regions at intermediate latitudes show positive residuals, which could suggest enhanced CR sources in the outer Galaxy, a hypothesis supported by gamma-ray data \citep{diffuse2,O2019} and radio data \citep{OS2013}. 
Hence, our analysis indicates that conventional parameter values, typically adopted in CR and gamma-ray studies, underestimate the synchrotron emission at intermediate latitudes in the outer Galaxy.
Another potential explanation is that the Fan region extends more widely than assumed, although this would not fully account for the observed enhancement in the entire anti-center regions. 
A more detailed quantitative assessment of the residuals would require extending this study to lower Galactic latitudes. This issue will need to be addressed in future studies of the Galactic disk magnetic field, which lies beyond the scope of the present work.

\paragraph{The Local Bubble.} 
For the Local Bubble to entirely explain the large-scale synchrotron excess without invoking an ordered random component in the halo, the required bubble parameters appear difficult to reconcile with an evolved structure such as the Local Bubble.
For example, for the 4~kpc propagation halo model, $B_\perp$ within a thin shell would need to be about 30--40 times larger than the average local value of the coherent disk field reported by \citet{Xu2024}, which is highly unlikely.
Hence, while a possible contribution from the Local Bubble cannot be excluded by our method, our conclusion regarding the existence of an ordered random magnetic-field component remains robust against this possibility. In other words, the Local Bubble alone cannot explain such a large discrepancy without invoking a large-scale ordered random component.
A geometrical model of the Local Bubble, such as those presented in \citet{Korochkin} and \citet{Pelgrims2025}, is not sufficient to explain the synchrotron excess we find. 
The differences between this work and \citet{Pelgrims2025} may be attributed to differences in the initial model assumptions, as well as in the modeling and fitting procedures.
The discrepancy between the conclusions of this work and those of  \citet{Korochkin} may arise in part from the different CR electron model adopted and primarily from the different coherent halo magnetic field employed.
Indeed, their large coherent halo component, which is a factor of 4.3 stronger than the value reported by \citet{Xu2024} and adopted in our work, may compensate for the absence of the ordered random component. In fact, the maximum strength of their coherent component is close to our derived ordered component, which also includes our best-fit ordered random contribution. The coherent magnetic-field estimate of \citet{Xu2024} is significantly smaller and is free of local contamination, which may explain the stronger coherent toroidal component used by \citet{Korochkin}. 
This may also explain the larger magnetic-field strengths derived in our analysis 
of about $50~\mu\mathrm{G}$ ($30~\mu\mathrm{G}$) in an extremely thin Local Bubble shell size of $\sim 5~\mathrm{pc}$ ($\sim 15~\mathrm{pc}$) for a bubble of $200~\mathrm{pc}$ ($500~\mathrm{pc}$) radius, compared to the values found by \citet{Korochkin}, \citet{Pelgrims} and others \citep[e.g.][]{Oneill,Maconi}, thereby disfavoring the interpretation in which the Local Bubble alone explains the synchrotron excess. Our result is also consistent with the recent finding of \citet{Susan} that there is no evidence for an imprint of the Local Bubble geometry on the dust polarization fraction. \\

In conclusion, we observe a significant large-scale synchrotron excess relative to the model predictions using the original coherent component strength reported by \citet{Xu2024}.
Our results favor an interpretation in which a large fraction of the synchrotron excess arises from an ordered random magnetic-field component in the halo, as a Local Bubble with standard parameters is unlikely to fully account for the observed emission, although non-standard  Local Bubble configurations cannot be excluded.
We find that, depending on the assumed CR propagation halo size, the total ordered toroidal magnetic field halo strength, modeled as the coherent field in \citet{Xu2024}, lies in the range \((3.2\text{--}4.3)\,\mu\mathrm{G}\), which is approximately a factor of 4--6 larger than the original strength of the coherent magnetic field inferred from rotation measures. This result provides compelling evidence for the presence of a significant ordered random magnetic field component in the Galactic halo, suggesting that such a component plays a fundamental role in shaping the observed polarized synchrotron emission.
In conclusion, within the framework of the adopted magnetic field and CRe spectrum and distributions, an additional ordered random magnetic field in the halo and a 4~kpc propagation halo provide a more accurate representation of the Galactic synchrotron sky. 
The resulting magnetic-field model and strength can be used in other studies, such as the deflection of extragalactic CR, the propagation of Galactic CR, foreground modeling for CMB studies, star and Galaxy formation, and for studies on external galaxies.

\begin{acknowledgements}

The {\it Planck} facility is acknowledged. Some of the results in this article have been derived using \texttt{HEALPix} software packages \citep{Healpix} available on the JPL NASA websites. This research used Astropy, a community-developed core Python package for Astronomy \citep{astropy2013, astropy2018}. The NASA/IPAC Infrared Science Archive, which is funded by the NASA and operated by the California Institute of Technology, is acknowledged for the Planck Public Data Release 3 Maps. 
We also acknowledge the use of the Legacy Archive for Microwave Background Data Analysis (LAMBDA) supported by the Astrophysics Science Division at NASA and the ASI Cosmic Ray Database at the Space Science Data Center. 
NASA Grant n. 80NSSC22K0495 is acknowledged.
\end{acknowledgements}

%

\begin{thebibliography}{}
\bibitem[Ackermann et al.(2012)]{diffuse2} Ackermann, M., Ajello, M., Atwood, W.~B., et al.\ 2012, \apj, 750, 1, 3
\bibitem[Aguilar et al.(2014)]{AMS02a} Aguilar, M., Aisa, D., Alvino, A., et al.\ 2014, \prl, 113, 12, 121102
\bibitem[Aguilar et al.(2021)]{AMS02} Aguilar, M., Ali Cavasonza, L., Ambrosi, G., et al.\ 2021, \physrep, 894, 1
\bibitem[Alves et al.(2018)]{Alves} Alves, M.~I.~R., Boulanger, F., Ferri{\`e}re, K., et al.\ 2018, \aap, 611, L5
\bibitem[Astropy Collaboration et al.(2013)]{astropy2013} Astropy Collaboration, Robitaille, T.~P., Tollerud, E.~J., et al.\ 2013, \aap, 558, A33
\bibitem[Astropy Collaboration et al.(2018)]{astropy2018} Astropy Collaboration, Price-Whelan, A.~M., Sip{\H{o}}cz, B.~M., et al.\ 2018, \aj, 156, 3, 123
\bibitem[Atwood et al.(2009)]{Atwood} Atwood, W.~B., Abdo, A.~A., Ackermann, M., et al.\ 2009, \apj, 697, 2, 1071
\bibitem[Beck et al.(1996)]{BeckRev1996} Beck, R., Brandenburg, A., Moss, D., et al.\ 1996, \araa, 34, 155
\bibitem[Beck et al.(2003)]{Beck2003} Beck, R., Shukurov, A., Sokoloff, D., et al.\ 2003, \aap, 411, 99
\bibitem[Beck(2015)]{BeckRev} Beck, R.\ 2015, \aapr, 24, 4
\bibitem[Bennett et al.(2013)]{Bennett} Bennett, C.~L., Larson, D., Weiland, J.~L., et al.\ 2013, \apjs, 208, 2, 20
\bibitem[BeyondPlanck Collaboration et al.(2023)]{BPColl2023} BeyondPlanck Collaboration, Andersen, K.~J., Aurlien, R., et al.\ 2023, \aap, 675, A1
\bibitem[Brandenburg et al.(1995)]{Brandenburg} Brandenburg, A., Moss, D., \& Shukurov, A.\ 1995, \mnras, 276, 651
\bibitem[Carretti et al.(2010)]{Carretti} Carretti, E., Haverkorn, M., McConnell, D., et al.\ 2010, \mnras, 405, 3, 1670
\bibitem[Clark \& Hensley(2019)]{Clark} Clark, S.~E. \& Hensley, B.~S.\ 2019, \apj, 887, 2, 136
\bibitem[Crutcher(2012)]{Crutcher2012} Crutcher, R.~M.\ 2012, \araa, 50, 29
\bibitem[Davis \& Greenstein(1951)]{Davis} Davis, L. \& Greenstein, J.~L.\ 1951, \apj, 114, 206
\bibitem[Dembinski et al.(2025)]{iminuit} Dembinski, H., Ongmongkolkul, P., Deil, C., et al.\ 2025, Zenodo, v2.31.1
\bibitem[de la Hoz et al.(2023)]{Quijote} de la Hoz, E., Barreiro, R.~B., Vielva, P., et al.\ 2023, \mnras, 519, 3, 3504
\bibitem[Dickinson et al.(2011)]{DickinsonDust} Dickinson, C., Peel, M., \& Vidal, M.\ 2011, \mnras, 418, 1, L35
\bibitem[Dickinson(2018)]{DickinsonLoops} Dickinson, C.\ 2018, Galaxies, 6, 2, 56
\bibitem[Ferri{\`e}re \& Terral(2014)]{FT2014} Ferri{\`e}re, K. \& Terral, P.\ 2014, \aap, 561, A100 
\bibitem[Ferri{\`e}re(2001)]{FerriereRev} Ferri{\`e}re, K.~M.\ 2001, Reviews of Modern Physics, 73, 4, 1031 
\bibitem[Ferri{\`e}re \& Schmitt(2000)]{FerriereDynamo} Ferri{\`e}re, K. \& Schmitt, D.\ 2000, \aap, 358, 125
\bibitem[Ferri{\`e}re(1994)]{FerriereNature} Ferri{\`e}re, K.\ 1994, \nat, 368, 6470, 403
\bibitem[Fauvet et al.(2011)]{Fauvet} Fauvet, L., Mac{\'\i}as-P{\'e}rez, J.~F., Aumont, J., et al.\ 2011, \aap, 526, A145
 \bibitem[Fauvet et al.(2012)]{Fauvet2012} Fauvet, L., Mac{\'\i}as-P{\'e}rez, J.~F., Jaffe, T.~R., et al.\ 2012, \aap, 540, A122
\bibitem[Guidi et al.(2023)]{Guidi} Guidi, F., G{\'e}nova-Santos, R.~T., Rubi{\~n}o-Mart{\'\i}n, J.~A., et al.\ 2023, \mnras, 519, 3, 3460
\bibitem[G{\'o}rski et al.(2005)]{Healpix} G{\'o}rski, K.~M., Hivon, E., Banday, A.~J., et al.\ 2005, \apj, 622, 2, 759
\bibitem[Halal et al.(2024)]{Susan} Halal, G., Clark, S.~E., \& Tahani, M.\ 2024, \apj, 973, 1, 54
\bibitem[Han et al.(1997)]{Han1997} Han, J.~L., Manchester, R.~N., Berkhuijsen, E.~M., et al.\ 1997, \aap, 322, 98. 
\bibitem[Han et al.(1999)]{Han1999} Han, J.~L., Manchester, R.~N., \& Qiao, G.~J.\ 1999, \mnras, 306, 2, 371
\bibitem[Han et al.(2004)]{Han2004} Han, J.~L., Ferriere, K., \& Manchester, R.~N.\ 2004, \apj, 610, 2, 820 
\bibitem[Han et al.(2006)]{Han2006} Han, J.~L., Manchester, R.~N., Lyne, A.~G., et al.\ 2006, \apj, 642, 2, 868
\bibitem[Han(2017)]{Han2017Rev} Han, J.~L.\ 2017, \araa, 55, 1, 111
\bibitem[Hanasz et al.(2009)]{Hanasz} Hanasz, M., W{\'o}lta{\'n}ski, D., \& Kowalik, K.\ 2009, \apjl, 706, 1, L155 
\bibitem[Haslam et al.(1982)]{Haslam} Haslam, C.~G.~T., Salter, C.~J., Stoffel, H., et al.\ 1982, \aaps, 47, 1
\bibitem[Haverkorn \& Heesen(2012)]{Haverkorn} Haverkorn, M. \& Heesen, V.\ 2012, \ssr, 166, 1-4, 133
\bibitem[Haverkorn et al.(2019)]{imagine} Haverkorn, M., Boulanger, F., En{\ss}lin, T., et al.\ 2019, Galaxies, 7, 1, 17
\bibitem[Heiles(2000)]{Heiles} Heiles, C.\ 2000, \aj, 119, 2, 923
\bibitem[Jaffe et al.(2010)]{Jaffe2010} Jaffe, T.~R., Leahy, J.~P., Banday, A.~J., et al.\ 2010, \mnras, 401, 2, 1013
\bibitem[Jaffe(2019)]{JaffeRev} Jaffe, T.~R.\ 2019, Galaxies, 7, 2, 52
\bibitem[Jaffe et al.(2013)]{Jaffe2013} Jaffe, T.~R., Ferri{\`e}re, K.~M., Banday, A.~J., et al.\ 2013, \mnras, 431, 1, 683
\bibitem[Jansson \& Farrar(2012)]{JF2012} Jansson, R. \& Farrar, G.~R.\ 2012, \apj, 757, 1, 14
\bibitem[Korochkin et al.(2025)]{Korochkin} Korochkin, A., Semikoz, D., \& Tinyakov, P.\ 2025, \aap, 693, A284
\bibitem[Maconi et al.(2025)]{Maconi} Maconi, E., Reissl, S., Soler, J.~D., et al.\ 2025, \aap, 698, A84
\bibitem[Manchester(1972)]{Manchester} Manchester, R.~N.\ 1972, \apj, 172, 43
\bibitem[Mao et al.(2012)]{Mao} Mao, S.~A., McClure-Griffiths, N.~M., Gaensler, B.~M., et al.\ 2012, \apj, 755, 1, 21
\bibitem[Martire et al.(2023)]{filaments} Martire, F.~A., Banday, A.~J., Mart{\'\i}nez-Gonz{\'a}lez, E., et al.\ 2023, \jcap, 2023, 4, 049
\bibitem[Mertsch \& Sarkar(2013)]{MS} Mertsch, P. \& Sarkar, S.\ 2013, \jcap, 2013, 6, 041
\bibitem[Miville-Desch{\^e}nes et al.(2008)]{Miville} Miville-Desch{\^e}nes, M.-A., Ysard, N., Lavabre, A., et al.\ 2008, \aap, 490, 3, 1093
\bibitem[O'Neill et al.(2024)]{Oneill} O'Neill, T.~J., Zucker, C., Goodman, A.~A., et al.\ 2024, \apj, 973, 2, 136
\bibitem[Orlando \& Strong(2013)]{OS2013} Orlando, E. \& Strong, A.\ 2013, \mnras, 436, 3, 2127
\bibitem[Orlando(2018)]{O2018} Orlando, E.\ 2018, \mnras, 475, 2, 2724
\bibitem[Orlando(2019)]{O2019} Orlando, E.\ 2019, \prd, 99, 4, 043007
\bibitem[Page et al.(2007)]{Page} Page, L., Hinshaw, G., Komatsu, E., et al.\ 2007, \apjs, 170, 2, 335 
\bibitem[Parker(1971)]{Parker71} Parker, E.~N.\ 1971, \apj, 163, 279
\bibitem[Panopoulou et al.(2025)]{Gina} Panopoulou, G.~V., Markopoulioti, L., Bouzelou, F., et al.\ 2025, \apjs, 276, 1, 15
\bibitem[Pelgrims et al.(2025)]{Pelgrims2025} Pelgrims, V., Unger, M., \& Mari{\textcommabelow s}, I.~C.\ 2025, \aap, 695, A148
\bibitem[Pelgrims et al.(2024)]{Pelgrims} Pelgrims, V., Mandarakas, N., Skalidis, R., et al.\ 2024, \aap, 684, A162
\bibitem[Pelgrims et al.(2020)]{Pelgrims2020} Pelgrims, V., Ferri{\`e}re, K., Boulanger, F., et al.\ 2020, \aap, 636, A17
\bibitem[Planck Collaboration et al.(2015)]{Planck2015a} Planck Collaboration, Ade, P.~A.~R., Aghanim, N., et al.\ 2015, \aap, 576, A104
\bibitem[Planck Collaboration et al.(2016)]{Planck2016} Planck Collaboration, Adam, R., Ade, P.~A.~R., et al.\ 2016, \aap, 594, A1
\bibitem[Planck Collaboration et al.(2016)]{PlanckBfield} Planck Collaboration, Adam, R., Ade, P.~A.~R., et al.\ 2016, \aap, 596, A103
\bibitem[Planck Collaboration et al.(2020)]{PlanckDR3} Planck Collaboration, Akrami, Y., Ashdown, M., et al.\ 2020, \aap, 641, A4 
\bibitem[Potgieter(2017)]{Potgieter} Potgieter, M.~S.\ 2017, Advances in Space Research, 60, 4, 848
\bibitem[Prouza \& {\v{S}}m{\'\i}da(2003)]{Prouza} Prouza, M. \& {\v{S}}m{\'\i}da, R.\ 2003, \aap, 410, 1
\bibitem[Pshirkov et al.(2011)]{Pshirkov} Pshirkov, M.~S., Tinyakov, P.~G., Kronberg, P.~P., et al.\ 2011, \apj, 738, 2, 192
\bibitem[Remazeilles et al.(2015)]{408MHz} Remazeilles, M., Dickinson, C., Banday, A.~J., et al.\ 2015, \mnras, 451, 4, 4311
\bibitem[Rybicki \& Lightman(1986)]{RL} Rybicki, G.~B. \& Lightman, A.~P.\ 1986, , 400
\bibitem[Taylor et al.(2009)]{Taylor} Taylor, A.~R., Stil, J.~M., \& Sunstrum, C.\ 2009, \apj, 702, 2, 1230
\bibitem[Terral \& Ferri{\`e}re(2017)]{TF2017} Terral, P. \& Ferri{\`e}re, K.\ 2017, \aap, 600, A29 
\bibitem[Tinyakov \& Tkachev(2002)]{TT} Tinyakov, P.~G. \& Tkachev, I.~I.\ 2002, Astroparticle Physics, 18, 2, 165
\bibitem[Silver \& Orlando(2024)]{SO2024} Silver, E. \& Orlando, E.\ 2024, \apj, 963, 2, 111
\bibitem[Stein et al.(2023)]{Stein} Stein, M., Heesen, V., Dettmar, R.-J., et al.\ 2023, \aap, 670, A158
\bibitem[Strong \& Moskalenko(1998)]{SM98} Strong, A.~W. \& Moskalenko, I.~V.\ 1998, \apj, 509, 1, 212
\bibitem[Strong et al.(2011)]{SOJ2011} Strong, A.~W., Orlando, E., \& Jaffe, T.~R.\ 2011, \aap, 534, A54
\bibitem[Strong et al.(2007)]{StrongRev} Strong, A.~W., Moskalenko, I.~V., \& Ptuskin, V.~S.\ 2007, Annual Review of Nuclear and Particle Science, 57, 1, 285
\bibitem[Sun et al.(2008)]{Sun2008} Sun, X.~H., Reich, W., Waelkens, A., et al.\ 2008, \aap, 477, 2, 573
\bibitem[Sun \& Reich(2010)]{Sun2010} Sun, X.-H. \& Reich, W.\ 2010, Research in Astronomy and Astrophysics, 10, 12, 1287
\bibitem[Svalheim et al.(2023)]{BPMap} Svalheim, T.~L., Andersen, K.~J., Aurlien, R., et al.\ 2023, \aap, 675, A14 
\bibitem[Unger \& Farrar(2024)]{UF2024} Unger, M. \& Farrar, G.~R.\ 2024, \apj, 970, 1, 95
\bibitem[Vidal et al.(2015)]{Vidal2015} Vidal, M., Dickinson, C., Davies, R.~D., et al.\ 2015, \mnras, 452, 1, 656
\bibitem[Xu \& Han(2024)]{Xu2024} Xu, J. \& Han, J.~L.\ 2024, \apj, 966, 2, 240
\bibitem[Xu \& Han(2019)]{Xu2019} Xu, J. \& Han, J.~L.\ 2019, \mnras, 486, 3, 4275
\bibitem[Waelkens et al.(2009)]{Hammurabi} Waelkens, A., Jaffe, T., Reinecke, M., et al.\ 2009, \aap, 495, 2, 697
\bibitem[Watts et al.(2024)]{Watts2024} Watts, D.~J., Fuskeland, U., Aurlien, R., et al.\ 2024, \aap, 686, A297. 
\bibitem[Watts et al.(2023)]{Watts2023a} Watts, D.~J., Galloway, M., Ihle, H.~T., et al.\ 2023, \aap, 675, A16

\end{thebibliography}

\end{document}